\documentclass[twocolumn,showpacs,preprintnumbers,amsmath,amssymb]{revtex4}
\usepackage{graphicx}
\usepackage{dcolumn}
\usepackage{bm}

\begin{document}
\title{Spin-polaron band structure and hole pockets
in underdoped cuprates}
\author{P. Wr\'obel$^1$, W. Suleja$^1$ and  R. Eder$^2$}
\affiliation{$^1$Institute for the Low Temperature and Structure
Research, P.O. Box 1410, 50-950 Wroc{\l}aw 2, Poland\\
and\\$^2$Forschungszentrum Karlsruhe, Institut f\"ur
Festk\"orperphysik, P.O.Box 3640, D-76021
Karlsruhe, Germany} 
\begin{abstract}
  We present a variational approach based on the string picture
  to analyze the internal structure and dispersion of spin polarons 
  with different symmetries in an antiferromagnet. We then use 
  this to discuss the properties of underdoped cuprate superconductor 
  within the `doped insulator' picture. The theory explains
  the remnant Fermi surface for the undoped compunds, as well as hole
  pockets, Fermi arcs, high energy pseudogap and the the mid-infrared 
  band in doped materials. Destructive interference between the phases 
  of a photohole near $\Gamma$ and the internal phases of the Zhang
  Rice singlet combined with our theory moreover explains the `waterfall' 
  phenomenon.
\end{abstract}

\maketitle
\section{Introduction}
Since the discovery of high temperature superconductivity (HTSC) in
doped antiferromagnetic (AF) insulators, the research on that
phenomenon concentrates to a great extent on the properties of
single-particle-like excitations in such systems. It is also obvious that the
microscopic mechanism of HTSC should hinge on the interaction between
quasiparticles.  
At half filling the Hubbard model, which is a generic model used to
describe strongly correlated systems, is an
insulator for large enough $U/t$\cite{Hubbard} whereas for low electron density the
model is expected to be a Fermi liquid with a Fermi surface volume
in accordance with the Luttinger theorem for any
$U/t$\cite{Kanamori}. With increasing
doping one might therefore expect a phase transition from a
`correlation dominated' phase near half-filling to a Fermi-liquid
phase for low density.  The key property of the correlation
dominated phase thereby is the splitting of the physical electron
into the two Hubbard bands which correspond to Fermionic holes and
double occupancies moving in a `background' of singly occupied sites,
whereby the electrons forming the background retain only
their spin degrees of freedom.  This - and not a half-filled Fermi
surface - is the picture underlying all successful theories for the
angle resolved photoemission spectroscopy (ARPES) data\cite{Wellsetal93}
obtained in insulating compounds
\cite{Trugman88,KyungFerrel,LemaAligia,Chernyshevetal94,Belinicheratal95,Balaetal95,Plakidaetal97,Belinicheretal97,Sushkovetal97}. 
The question then is to
what extent these Hubbard bands can be doped before the two-band
structure collapses, the spin background `melts' into a Fermi sea and
the Luttinger Fermi surface is regained.  The key property of this
`doped insulator' phase should be a Fermi surface with a volume
proportional to the number of doped holes because these are the only
mobile Fermions.

ARPES on cuprate superconductors
has produced a wealth of information\cite{Damareview}
although it has proven
difficult to extract a consistent picture. From spectra taken
near optimal doping it was concluded that
ARPES shows evidence for a `large' Fermi surface consistent
with the Luttinger theorem and band calculations\cite{Damareview}
which would imply
that the doped Hubbard bands never exist. Surprisingly enough, however, 
insulating compounds such as Sr$_2$CuO$_2$Cl$_2$ seemed to show  a 
very similar Fermi surface as well - which has been termed the 
`remnant Fermi surface'\cite{Ronningetal98}.  ARPES spectra from insulators
show a band which disperses towards lower binding
energy and then rapidly looses weight
just as if it would cross a Fermi level.
The only possible explanation for this phenomenon
is a strong and systematic variation of the spectral weight
of the conduction band which drops to near zero abruptly
at a line in ${\bf k}$-space which roughly coincides with the
noninteracting Fermi surface. One can give simple arguments
why such a behaviour is to be expected\cite{comment}. 
Since there is no reason why such
a strong variation of spectral weight and bands with almost no
spectral weight should occur only in insulators one should be cautioned
that Fermi surface maps for doped compounds may not show the 
full picture either.

In the underdoped compounds the upper Hubbard band, while rapidly
loosing spectral weight, still can be clearly resolved\cite{Chenetal92}.
ARPES shows rather structureless
spectra which are usually interpreted in terms of a `high energy
feature' and a `leading edge shift' or, alternatively,
a high energy and low energy pseudogap\cite{Damareview}. 
In any case there is definitely
no `large' Fermi surface, instead Fermi surface maps show `Fermi
arcs'. Bearing in mind the remnant Fermi surface
in the insulators 
suggests to interpret these arcs as being the inner part
of a hole pocket centered on $(\frac{\pi}{2},\frac{\pi}{2})$
with the part of the pocket facing $(\pi,\pi)$ having too small
spectral weight to be seen in ARPES. Assuming that the
electronic structure in the underdoped compound
can be described in simplest approximation
as holes doped into the quasiparticle band
of the insulator moreover would give an 
explanation for the high energy pseudogap - the dispersion
of the `high energy feature' - in that it simply reflects the
hole dispersion in the insulator\cite{Damareview}. Such a `strict' rigid band
picture would not explain the low energy pseudogap
or the temperature and doping dependence of either high or low energy 
pseudogap. On the other hand the closing of the pseudogap with
increasing doping can be seen already in cluster
simulations\cite{EOS} and it has been pointed out there
that the mechanism may be an effective downward renormalization
of the $t'$ and $t''$ terms due to decrease in the spin correlation
length. Moreover, the low energy pseudogap
by its definition in terms of the leading edge shift
has no immediate connection with a dispersion
and may be determined e.g. by $T$-dependent linewidths 
as discussed by Storey {\em et al.}\cite{Storeyetal2007}.

The compound Ca$_{2-x}$Na$_x$CuO$_2$Cl$_2$
shows rather clear evidence for
the doped insulator picture\cite{Kohsakaetal02} in that the dispersion 
in the doped case is virtually identical to that in the undoped compound
and even the part of the quasiparticle band facing $(\pi,\pi)$
may have been observed.Very recently ARPES
experiments seem to have provided direct evidence for
hole pockets in La$_{1.48}$Nd$_{0.4}$Sr$_{0.12}$CuO$_4$\cite{Changetal},
with the part of the pockets facing $(\pi,\pi)$ indeed having small
spectral weight - as one would have expected on the basis of the
remnant Fermi surface.

It had been noted early on \cite{Trugman} that hole pockets with a volume
proportional to the hole concentration would 
explain the scaling of the low temperature
Hall constant with hole concentration in the 
underdoped materials \cite{Ongetal87,Takagietal89}
and that the apparent discrepancy between the `small Fermi surface'
suggested by transport measurements and the `large Fermi surface'
seen in ARPES may be due to a systematic variation
of the quasiparticle weight along the hole pocket.\cite{rigid}.
While the recent discovery \cite{Doiron-Leyraudetal07,Yelland,Bangura} of 
Shubnikov-deHaas oscillations in
some underdoped compounds initially seemed to provide strong
evidence for hole pockets the finding that the oscillations
are likely caused by electron-like rather than hole-like 
pockets\cite{LeBoeuf}
has complicated matters again. 

Lastly, exact diagonalization studies of the t-J model
provide clear evidence that the Fermi surface
for hole dopings around 10$\%$ takes the form of
hole pockets\cite{pockets,Leung}.
Careful analysis of exact diagonalization results
shows that
the single particle the spectral function for the doped t-J model
is quite consistent with rigid band filling of the quasiparticle band
seen at half-filling, provided one takes into 
account the formation of hole pairs\cite{rigid}.
By calculating the spectral function for
dressed hole operators\cite{DagottoSchrieffer}
rather than the bare electron operators
is can moreover be shown that the quasiparticles in the doped system 
have very nearly the same internal structure as in the 
undoped one\cite{bags}. 

Here we take the point of view that the underdoped regime
in high-temperature superconductors precisely
corresponds to the doped insulator
phase. We show that many properties of the underdoped
phase - the remnant Fermi surface, the Fermi arcs,
the high-energy feature seen in ARPES, the mid infrared band seen
in the optical conductivity - find a simple and natural explanation in
the dispersion and internal structure of the quasiparticles which 
correspond to holes heavily dressed by spin excitations.
Since there are indications that the heavily overdoped phase
is essentially a Fermi liquid this would imply that the
phase transition from the correlation dominated phase to the
Fermi liquid phase occurs at optimal doping. This would then
be a quantum phase transition where none of the two phases
has any kind of order - rather they differ in the topology and volume
of their Fermi surfaces. An indication of this transition can in fact be
seen in the dynamical spin and density correlation function obtained
by  exact diagonalization of small clusters. In the underdoped regime
spin and density correlation functions are very different
and  the density correlation function takes
the form of extended incoherent continua\cite{anomalous}
This form of the density correlation function can be explained
quantitatively within the string picture for a 
single hole\cite{VojtaBecker97}.
For doping levels higher than optimal, spin and density correlation
function become similar and can be explained well
as particle-hole transitions across an essentially free-electron-like
Fermi surface\cite{intermediate}. 

While a theory for such a transition would be highly desirable
but very challenging the present paper has a more modest goal:
we want to show that many features of undoped and
underdoped cuprates can be explained by a very simple theory
which assumes continuity with the insulator.
The calculation will be performed in the framework of the t-J
model\cite{Chaoetal78,ZhangRice88}
extended by terms enabling hopping to second and third nearest
neighbors with hopping integrals $t'$ and $t''$ 
respectively\cite{Nazarenkoetal,Leungetal}. We use
standard values $t=0.35$eV, $t'=-0.12$ eV, $t''=0.08$ eV, and
$J=0.14$eV chosen so as to reproduce the measured Fermi surface of
hole doped cuprates for high doping levels.

\section{Construction of localized basis states}
Since we want to study the doped insulator we 
consider the motion of a single hole in an
antiferromagnetically ordered `spin background'.  All processes
analyzed in the following actually require only short range
antiferromagnetic correlations - one may therefore expect that hole
motion in a state with short range antiferromagnetic order but no long
range order will involve very similar processes so that e.g. the
internal structure of the quasiparticles and the
dispersion relation of a hole should not change drastically.  The
construction of spin polaron (SP) states including the excited
states was performed in several earlier publications
\cite{EderBecker90,EderBecker91,EderWrobel93,Ederetal96,WrobelEder98}.
In order to make this paper self-contained we will now briefly repeat
that construction.  For definiteness we will assume that a
$\downarrow$-spin has been removed from the system and study the
motion of the resulting hole.  We denote the $\downarrow$-sublattice
by $A$.

To begin with we define $H_0=H_t + H_{Ising}$ to be the sum of the 
nearest neighbor hopping $\propto t$
and the longitudinal part of the Heisenberg exchange:
\begin{eqnarray*}
H_t &=& -t\sum_{\langle i,j \rangle,\sigma}(\hat{c}_{i\sigma}^\dagger
\hat{c}_{j\sigma}^{} + H.c.), \nonumber
\\
H_{Ising} &=& J \sum_{\langle i,j \rangle} 
(S_i^z S_j^z - \frac{n_i n_j}{4}).
\end{eqnarray*}
$H_0$ is frequently referred to as the $t-J_z$ model.
In a first step, we seek approximate eigenstates of $H_0$ which are localized
due to the string effect.

\begin{figure}
\includegraphics[width=\columnwidth]{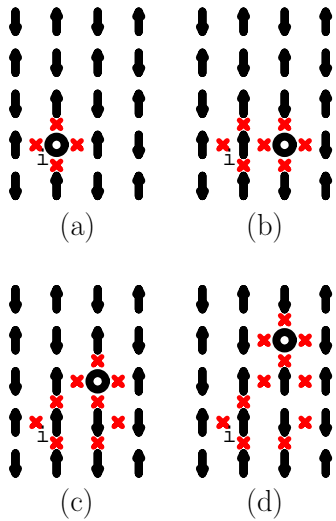}
\caption{\label{fig1}  (Color online) The mechanism of the string effect. Slanted
crosses represent links with the contribution to the Ising part of the
exchange energy higher by $J/2$ as compared to
the N\'eel state. (b), a string state obtained
by a single move of the hole created at site $i$. (c), (d), states
obtained respectively by two and three consecutive moves.}
\end{figure}

The mechanism of the string effect is shown in Fig.\ref{fig1}.
By creating a hole in the N\'eel state - i.e. the ground state
of $H_{Ising}$ - 
at site $i$ and acting repeatedly
with the hopping term we generate a basis of string states
$|{\cal P}_i\rangle$ where
${\cal P}_i= (i,j,\ldots,n)$ is shorthand for
the sites $i,j,\ldots,n$ visited by the hole. All spins on these sites
have been displaced by one lattice spacing and thus are
inverted relative to the N\'eel order.

All $|{\cal P}_i\rangle$ are eigenstates of $H_{Ising}$.
Taking the energy of the string with length $0$ i.e. the bare hole 
as the zero of energy and denoting by $\nu$ the length of
the string - i.e. the number of shifted spins - the eigenvalue is
\begin{equation}
E_\nu = \frac{J}{2} (2\;\nu +1)
\label{ising}
\end{equation}
This is exact for $\nu\le 2$ and is true for `most' longer
strings as well.
For simplicity we assume (\ref{ising}) to be true for any string. This
implies that the hole is trapped in a linearly ascending potential
and all eigenstates of $H_0$ are localized. The main deviations from
(\ref{ising}) occur for `loops' as discussed
by Trugman\cite{Trugman88} which in fact
lead to hole-propagation even in the $t-J_z$ model.
Such loops pose no fundamental problem
for the present formalism and can be dealt with by introducing 
the concept of `irreducible paths' as discussed in detail in
 Ref. \cite{EderBecker90}.

\begin{figure}
\includegraphics[width=\columnwidth]{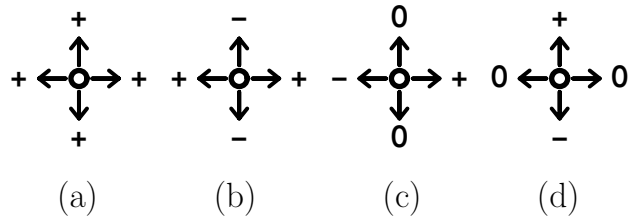}
\caption{\label{fig2} Dependence of the sign of a given path on the
direction of the first move for the $d_{x^2-y^2}$-wave SP, (b), $p_x$-wave SP,
(c), and $p_y$-wave SP, (d). For completeness the schematic
representation for the $s$-wave SP has also been shown, (a).}  
\end{figure}

Next we note that by acting with a point group operation, which leaves the
initial site $i$ invariant, the string states $|{\cal P}_i\rangle$
are transformed into one another. We can therefore
define linear combinations of the $|{\cal P}_i\rangle$
which transform like the basis states
of the irreducible representations of $C_{4v}$ under these
point group operations. Then we make the following
ansatz for a localized eigenstate of $H_0$ 
\begin{equation}
|\Psi^{(o,m)}_i\rangle = \sum_{{\cal P}_i}
\alpha^{(o,m)}_{{\cal P}_i} |{\cal P}_i\rangle,
\label{wvfn}
\end{equation}
where $o\in\{s,p_x,p_y,d_{x^2-y^2},d_{xy}\dots\}$ denotes the
symmetry or `orbital character'
of the state and $m$ labels the excitation number for
a given symmetry. 
In keeping with (\ref{ising}) we  moreover assume that
each coefficient $\alpha^{(o,m)}_{{\cal P}_i}$
can be factorized into a sign $\phi_{{\cal P}_i}^{(o)}$ 
- which plays the role of an
`angular wave function' - and a `radial wave function'
$\alpha^{(o,m)}_\nu$ which depends only
on the length of the string:
\begin{equation}
 \alpha^{(o,m)}_{{\cal P}_i} = \phi_{{\cal P}_i}\;
\alpha^{(o,m)}_\nu
\label{prefs}
\end{equation}
For an $A_1$ ($s$-wave) state the sign $\phi_{{\cal P}_i}$ is obviously
uniform for all paths. For an $E$ ($p$-wave) or
$B_1$ ($d_{x^2-y^2}$-wave) state  $\phi_{{\cal P}_i}$ is determined
by the direction of the first hop away from the site
$i$ as shown in Figure \ref{fig2}.
The string of length $0$ i.e. the bare hole at site
$i$ is invariant under all point group operations and
hence has nonvanishing weight only in the $s$-like state - this
implies that $E$ and $B_1$ states are higher in energy
than the $A_1$ state because the are composed
of strings with length $\ge 1$ and hence
a minimum of three frustrated
bonds. For the two remaining representations $A_2$ ($g$-wave) and
$B_2$ ($d_{xy}$-wave) it can be shown that
only strings with a minimum length of $2$ have nonvanishing weight
in the corresponding SP states - these states therefore
are even higher in energy and we omit them.
For each symmetry sector $o$ we can now set up the Schr\"odinger equation
\begin{equation}
H_0 |\Psi^{(o,m)}_i\rangle = E^{(o,m)} |\Psi^{(o,m)}_i\rangle.
\label{schroedinger}
\end{equation}
thereby assuming (\ref{ising})
and solve for the eigenenergies $E^{(o,m)}$ and the coefficients
$\alpha$ - this is explained in Appendix \ref{appa}. 
As expected for a linearly ascending potential,
the $\alpha$'s are rapidly decreasing with
increasing length of the string even for the physical parameter range
$t/J\approx 3$. It turns out the Schr\"odinger equation
for the coefficients
$\alpha$ for the $d_{x^2-y^2}$ SP is identical to that for the
$p$-like ones. To shorten the notation we call the
coefficients $\alpha^{(s,0)}_{\nu}$ for the lowest $s$-like SP
$\alpha_\nu$ and instead of those for the $p$-like SP,
$\alpha^{(p,0)}_{\nu}$ and $d_{x2-y2}$-like SP, $\alpha^{(d,0)}_{\nu}$
we use  $\alpha'_{\nu}/\sqrt{2}$ and $\alpha'_{\nu}/2$  respectively.

\section{Effective multi-band model for spin polarons}
\begin{figure}
\includegraphics[width=\columnwidth]{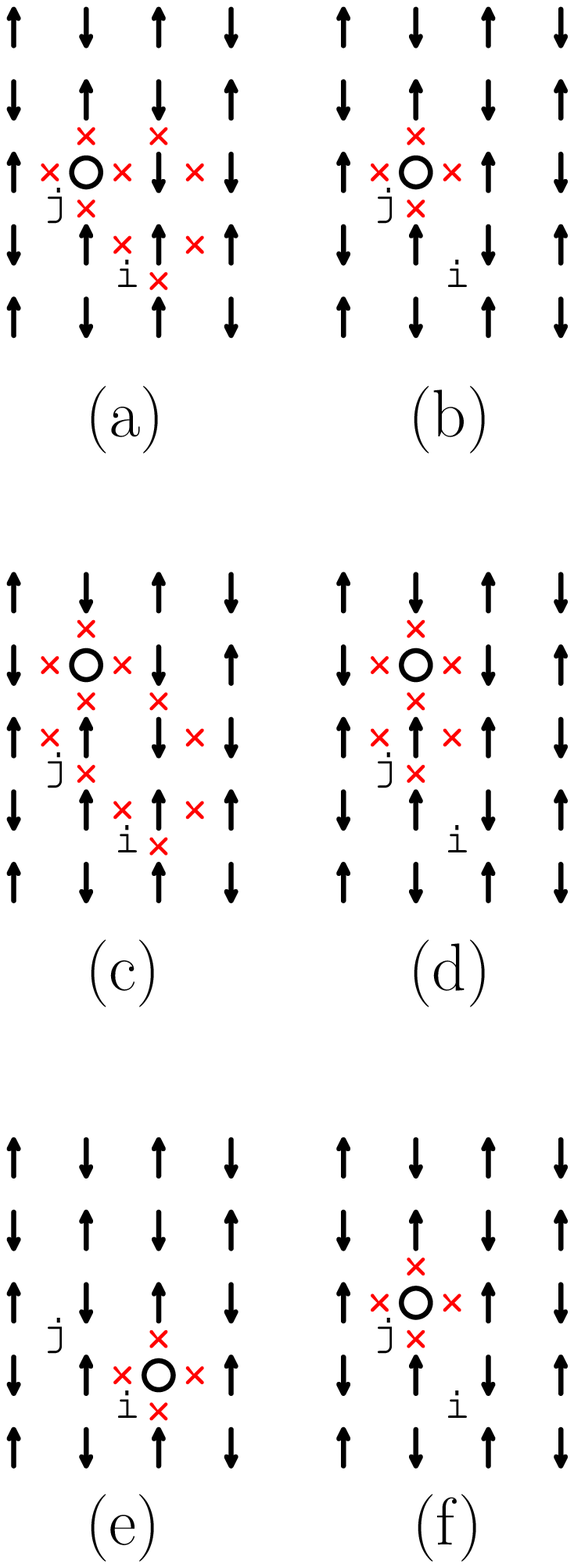}
\caption{\label{fig3} (Color online)
(a)$\rightarrow $(b): a string of length $2$ starting at $i$
is truncated to  a string of length $0$ at $j$ by the 
transverse part of the Heisenberg
exchange.\\
(c) $\rightarrow$ (d) a string of length $3$ is reduced to one of
length $1$.\\
(e) $\rightarrow$ (f) a bare hole at site $i$ (`string of length
$0$')
is transported to $j$ by $t'$-hopping.}

\end{figure}

So far we have found SP states (\ref{wvfn}) which form a set of
approximate localized eigenstates of 
$H_0$ at each of the sites of the sublattice $A$. 
Next we note that the remaining part $H_1$ of the
 $tJ$M - which comprises the transverse part of the Heisenberg exchange and
the hopping terms $\propto t',t''$ -
has nonvanishing matrix elements between SP states
centered on neighboring sites $i$ and $j$. One important
mechanism leading to such a matrix element is the truncation
of the string shown in Figure  \ref{fig3}.
By flipping the first two spins of the defect string the
starting point of the string is shifted to a second or third
nearest neighbor while the length of the string is reduced
by two. Since the coefficients $\alpha^{(o,m)}_{{\cal P}_i}$
and $\alpha^{(o')}_{{\cal P}'_j}$
of the initial and final string are known by solution of
(\ref{schroedinger}) and the strength of the spin-flip
term is $J/2$ the corresponding matrix element is easily
evaluated. Similarly, the hopping terms $\propto t',t''$ allow for 
the hopping of the bare hole between the sites of one sublattice
(see Figure \ref{fig3}).
In addition there is the `loop hopping'\cite{Trugman88} and actually a wide
variety of additional processes, which are discussed in 
Appendix \ref{appa}.

Assuming that the matrix elements are known
we define Fourier transforms
\begin{equation}
|\Psi^{(\lambda)}({\bf k})\rangle = \sqrt{\frac{2}{N}}
\sum_{j\in A}
|\Psi^{(\lambda)}_j\rangle e^{-i{\bf k} \cdot {\bf R}_j}
\label{bloch}
\end{equation}
where $\lambda=(o,m)$. Next we
make the LCAO-like ansatz for a propagating
single-hole state
\begin{equation}
|\Phi_l({\bf k})\rangle = \sum_\lambda v^{(l)}_{\lambda}({\bf k})
|\Psi^{(\lambda)}({\bf k})\rangle
\label{lcao_like}
\end{equation}
which leads to a generalized
eigenvalue problem of the form
\begin{equation}
H_{eff}({\bf k}) {\bf v}_l({\bf k}) = E_l({\bf k}) 
O_{eff}({\bf k}) {\bf v}_l({\bf k})
\label{eigvalprob}
\end{equation}
where the Hamilton and overlap matrices are given by
\begin{eqnarray}
H_{(\lambda \lambda')} &=& \delta_{\lambda \lambda'} E^{(\lambda)} 
+ {\cal T}^{(\lambda \lambda')}({\bf k}), \label{hmatelksp}  \\
O_{(\lambda \lambda')} &=& \delta_{\lambda \lambda'}  + {\cal
  O}^{(\lambda \lambda')}({\bf k}) 
 \label{ovmatelksp}
\end{eqnarray}
and we have introduced the Fourier transform of 
\begin{eqnarray}
{\cal T}^{(\lambda \lambda')}_{ii'}=\langle\Psi^{(\lambda)}_{i}|H_1|\Psi^{(\lambda')}_{i'}\rangle.
\label{hmatel}\\
{\cal O}^{(\lambda \lambda')}_{ii'}=\langle\Psi^{(\lambda)}_{i}|\Psi^{(\lambda')}_{i'}\rangle.
\label{ovmatel}
\end{eqnarray}
All these matrix elements can be expressed in terms of the
coefficients $\alpha^{(\lambda)}_l$ as discussed in the Appendix  \ref{appa}.
Once these matrix elements are calculated we readily obtain the
band structure for the spin polarons. 

To conclude this section we briefly discuss the relationship with
previous work. Several authors have studied hole motion in an
antiferromagnet by calculations within a string 
basis\cite{Trugman88,InoueMaekawa90,VojtaBecker98,Boncaetal07}.
The difference is that the diagonalization of
$H_0$ leads to a considerable reduction of basis states
in that high lying eigenstate of $H_0$ are eliminated from the very
beginning. The matrices to be diagonalized in the present
work are $4\times 4$ or $10\times 10$ - which is very small
compared to the matrix dimensions in 
Refs. \cite{Trugman88,InoueMaekawa90,VojtaBecker96,VojtaBecker98,Boncaetal07}.
Moreover the LCAO-like scheme makes it easier to
extract a physical picture. Another frequently applied approach is
the self-consistent Born 
approximation\cite{MartinezHorsch91,LiuManousakis92,Sushkovetal97,Chernyshevetal94,Belinicheratal95,Balaetal95,Plakidaetal97,Belinicheretal97,Manousakis07}.
The single-hole wave function associated with this approximation
actually also can be interpreted as a superposition of 
string states once the Fourier-transformed version of the wave function
in Ref. \cite{Reiter} is converted into real space. This explains
why the results e.g. for the dispersion of a single
hole are practically identical.

\section{Photoemission spectra and Fermi surface}

ARPES gives the information on the one-electron removal part of the
spectral function defined, at $T=0$, as
\begin{equation}
A^{-}({\bf k}, \omega)=
-\frac{1}{\pi} Im \langle \Psi_{AF} |c^{\dagger}_{{\bf k},\downarrow}
\frac{1}{\omega-H+i 0^{+}}  c_{{\bf k},\downarrow} | \Psi_{AF}  \rangle.
\label{sf}
\end{equation}
$| \Psi_{AF} \rangle$ represents in (\ref{sf}) the half-filled
ground state in which
the photoemission process takes place. Since we assume that
the rigid band scenario is applicable to cuprates in the low 
doping range we expect that the conclusions drawn from that analysis
are also to some extent applicable to doped systems.

\begin{figure}
\includegraphics[width=\columnwidth]{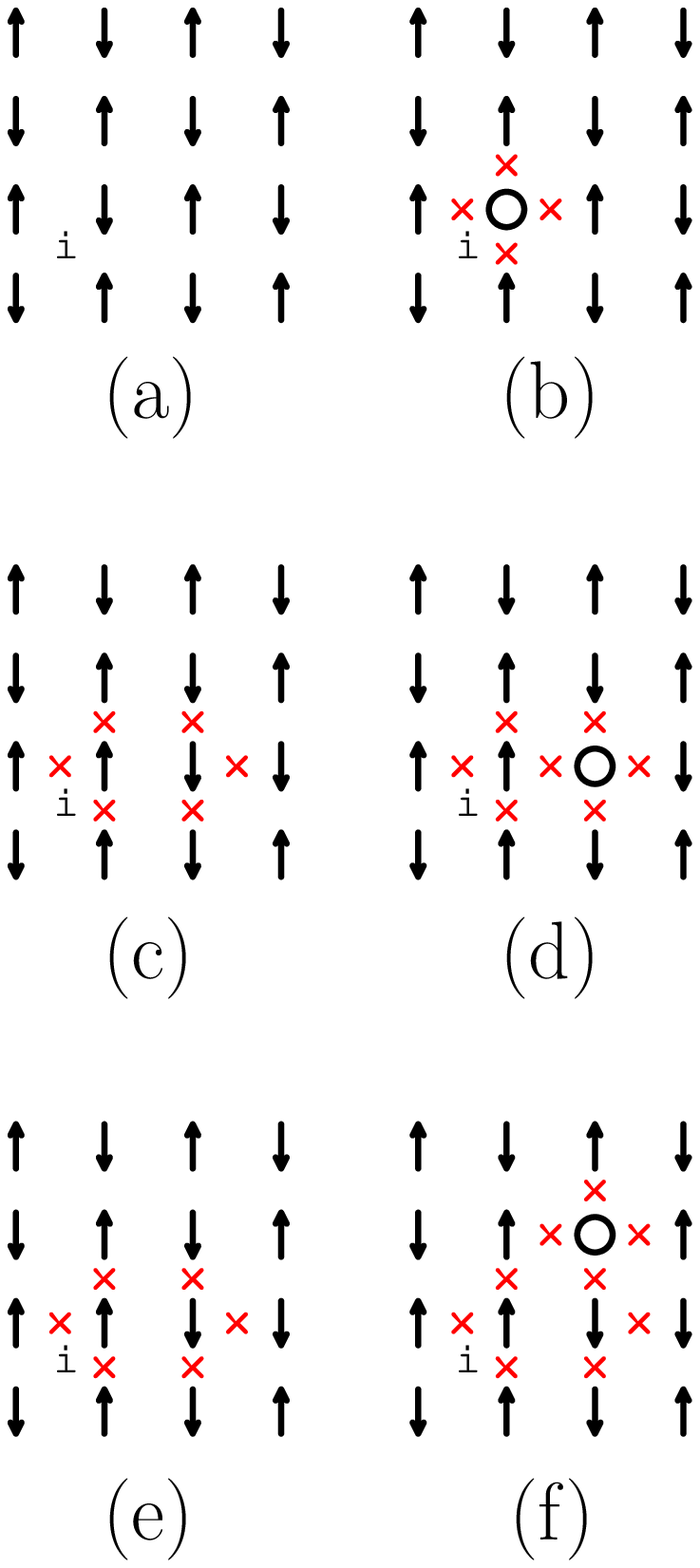}
\caption{\label{fig4} 
(Color online) (a) $\rightarrow$ (b):
Creation of a bare hole at site $i$ from the N\'eel state.
(c) $\rightarrow$ (d): Creation of a string of length $1$
starting at $i$ from the N\'eel state + quantum fluctuation.
(e) $\rightarrow$ (f): Creation of a string of length $2$
starting at $i$ from the N\'eel state + quantum fluctuation.}
\end{figure}

As a first step we approximate the resolvent operator 
$((\omega-H+i 0^{+})^{-1}$ by
\begin{equation}
\frac{1}{\omega-H+i 0^{+}} \to \sum_{l,{\bf k}} 
\frac{|\Phi_l({\bf k}) \rangle \langle
\Phi_l({\bf k})|}{\omega-E_l({\bf k})+i
0^{+}}.
\label{repl}
\end{equation}
i.e. we restrict the single hole states to the
coherent superposition of SP states (\ref{lcao_like}).\\
Next, we have to choose an approximate ground state $| \Psi_{AF} \rangle$ of
the Heisenberg antiferromagnet. The simplest choice would be
the N\'eel state $|\Phi_N\rangle$ but in this way we would miss an
important mechanism for a ${\bf k}$-dependent quasiparticle
weight, namely the coupling of the photohole to quantum
spin fluctuations.
By generating a hole in the N\'eel state we can obtain only 
a bare hole, i.e. the string of length $0$.
In the presence of quantum spin fluctuations the photoemission
process can also generate strings of length $1$ or $2$
- see Figure \ref{fig4}. Since in such a process the hole
is created not at the central site $i$ of the SP state -
which determines the phase factor $e^{-i {\bf k}\cdot {\bf R}_i}$
in the Bloch state (\ref{bloch}) - 
the photoemission matrix element becomes ${\bf k}$-dependent.
Generally speaking the fact that the SP quasiparticles
extend over more than one unit cell in real space results in
a `structure factor' which varies within the
first Brillouin zone in ${\bf k}$-space. As will be seen below, this very
${\bf k}$-dependence is the source of the 
remnant Fermi surface.
In order to capture this effect we use simple
first order perturbation theory for the quantum spin fluctuations
and set
\begin{equation}
| \Psi_{AF}  \rangle = |\Phi_N\rangle 
- \frac{1}{3}\sum_{\langle i,j\rangle} (S_i^+ S_j^- + H.c.)|\Phi_N\rangle .
\label{GS}
\end{equation}
Since second order perturbation theory gives quite a
good estimate for the ground state energy of the Heisenberg
antiferromagnet we expect that the probability for coupling to
a quantum fluctuation with the electron annihilation
operator is described quite well by (\ref{GS}). 

Since a quantum fluctuation in the
initial states simply gives rise to an extra
factor of $-1/3$ we immediately obtain the following
expressions for the photoemission matrix element
$m^{(\lambda)}=\langle\Psi^{(\lambda)}_{\downarrow}({\bf k})|
c_{{\bf k},\downarrow}| \Psi_{AF}  \rangle$
\begin{eqnarray}
m^{(s)}({\bf k}) &=& \alpha_0 -
\frac{2\alpha_1}{3} \left(\cos(k_x)+\cos(k_y)\right)\nonumber \\
&& 
- \frac{4\alpha_2}{3} \left(\left(\cos(k_x)+\cos(k_y)\right)2-1\right)\nonumber\\
m^{(p_x)}({\bf k}) &=&
\frac{\sqrt{2} i\alpha_1'}{3}\sin(k_x)\nonumber \\
&&
+ \frac{2 \sqrt{2} i\alpha_2'}{3}
\sin(k_x)\left(\cos(k_x)+\cos(k_y)\right)\nonumber\\
m^{(d)}({\bf k}) &=&
-\frac{\alpha_1'}{3} \left(\cos(k_x)-\cos(k_y)\right)\nonumber \\
&&
- \frac{\alpha_2'}{3} \left( \cos2(k_x)-\cos2(k_y)\right)
\label{pesmat}
\end{eqnarray}
Using these matrix elements we can now compute the spectral density
from the normalized SP eigenfunctions. 

If we want to compare to experiment, however, there is yet another
important effect we need to take into account, namely the
coupling of the photohole to charge fluctuations.
We may expect that the ground state of the
system has not only quantum spin fluctuations but also
charge fluctuations i.e. an admixture of pairs of holes
and double occupancies with a density $\propto (t/U)^2$.
By annihilating an electron on a doubly occupied site
it is possible to create a string state with an
initial site $i$ by annihilating an electron at a site
different from $i$. Again, this will give rise to a 
${\bf k}$-dependence of the spectral weight. 
Such processes actually are not described by the $tJ$M, but
since it is rather easy to discuss them we do so.
We treat the charge fluctuations in perturbation theory
i.e. we replace
\begin{equation}
|\Psi_{AF}\rangle \rightarrow
|\Psi_{AF}\rangle + \frac{t}{U}\sum_{\langle ij\rangle}
\sum_\sigma \hat{d}_{i\sigma}^\dagger \hat{c}_{j\sigma}^{}|\Phi_N\rangle 
\end{equation}
where $\hat{d}_{i\sigma}^\dagger =c_{i\sigma}^\dagger
n_{i\bar{\sigma}}$. The hopping terms $\propto t', t''$ do not
produce charge fluctuations in the N\'eel state.
Denoting $\eta=t/U=J/(4t)$ we get the following correction
to the matrix element $m^{(s)}$
\begin{eqnarray}
m^{(s)}({\bf k}) &=& 2 \eta \alpha_0 \left(\cos(k_x)+\cos(k_y)\right)
- 4 \eta \alpha_1
\label{pesmat1}
\end{eqnarray}
whereas the corrections to $m^{(p_x)}$ and $m^{(d)}$ are zero.
It might appear that the corrections due to charge fluctuations are
quite small, being anyway $\propto t/U$. On the other hand, by
coupling to charge fluctuations a bare hole
at site $i$ can be created by actually annihilating an electron
on any of its $z$ neighbors. By changing from ${\bf k}=(0,0)$
to ${\bf k}=(\pi,\pi)$ the corresponding
contribution to the photoemission matrix element
thus changes from $zt/U$ to $-zt/U$ and since this
has to be added before squaring the matrix element the impact
of the charge fluctuations is in fact quite strong.
The way in which we are treating charge fluctuations would be
adequate for a simple one-band Hubbard model, which is not the proper
model for cuprate superconductors. A very similar calculation for the
more correct two-band model has bee done by 
Eroles {\em et al.}\cite{Eroles}.

\begin{figure}
\includegraphics[width=\columnwidth]{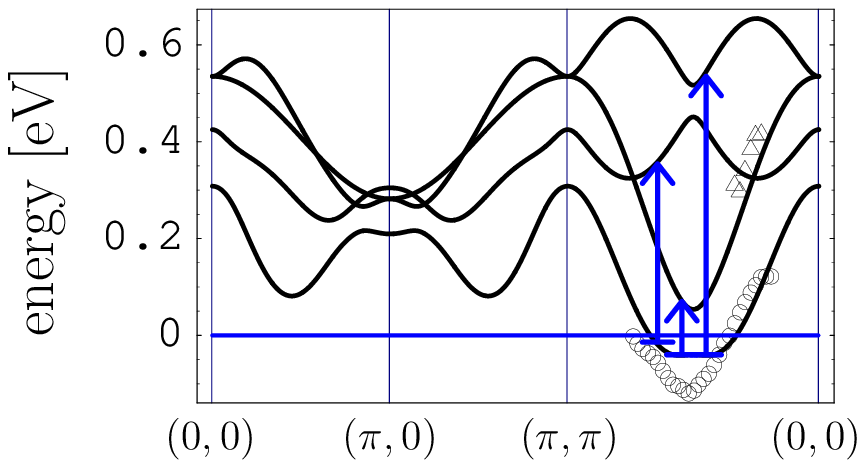}
\caption{\label{fig5} 
Band structure obtained by solving the eigenvalue problem
(\ref{eigvalprob}). Also shown is the dispersion of bands
observed by Ronning {\em at al.}in
Ca$_2$Cu0$_2$Cl$_2$\cite{Ronningetal05} (circles and triangles).
 A sample Fermi level of a doped system has been
chosen as the zero of energy. Vertical arrows label possible optical
transitions of a hole at the bottom of the band. }
\end{figure}

\begin{figure}
\includegraphics[width=\columnwidth]{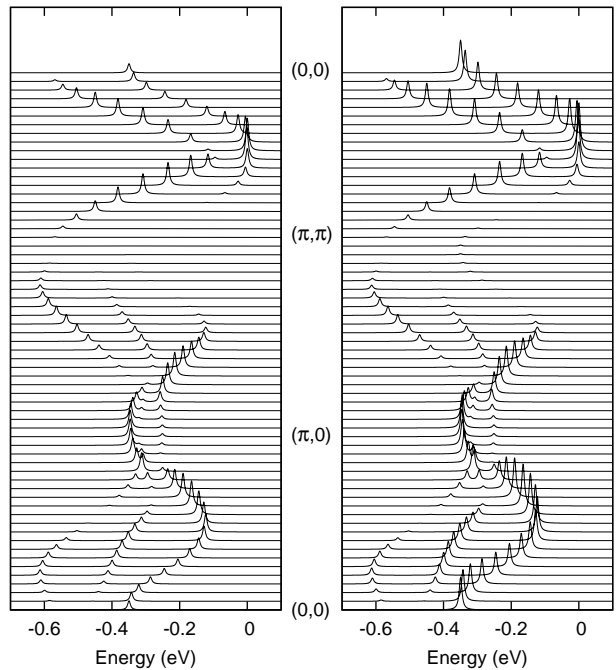}
\caption{\label{fig6} 
Photoemission spectrum for the half-filled
model corresponding to the band structure (\ref{fig5}).
In the left panel the spectral weight is computed using only the quantum
fluctuation correction (\ref{pesmat}), in the right panel
the charge-fluctuation correction (\ref{pesmat1}) was used as well.}
\end{figure}

\begin{figure}
\includegraphics[width=\columnwidth]{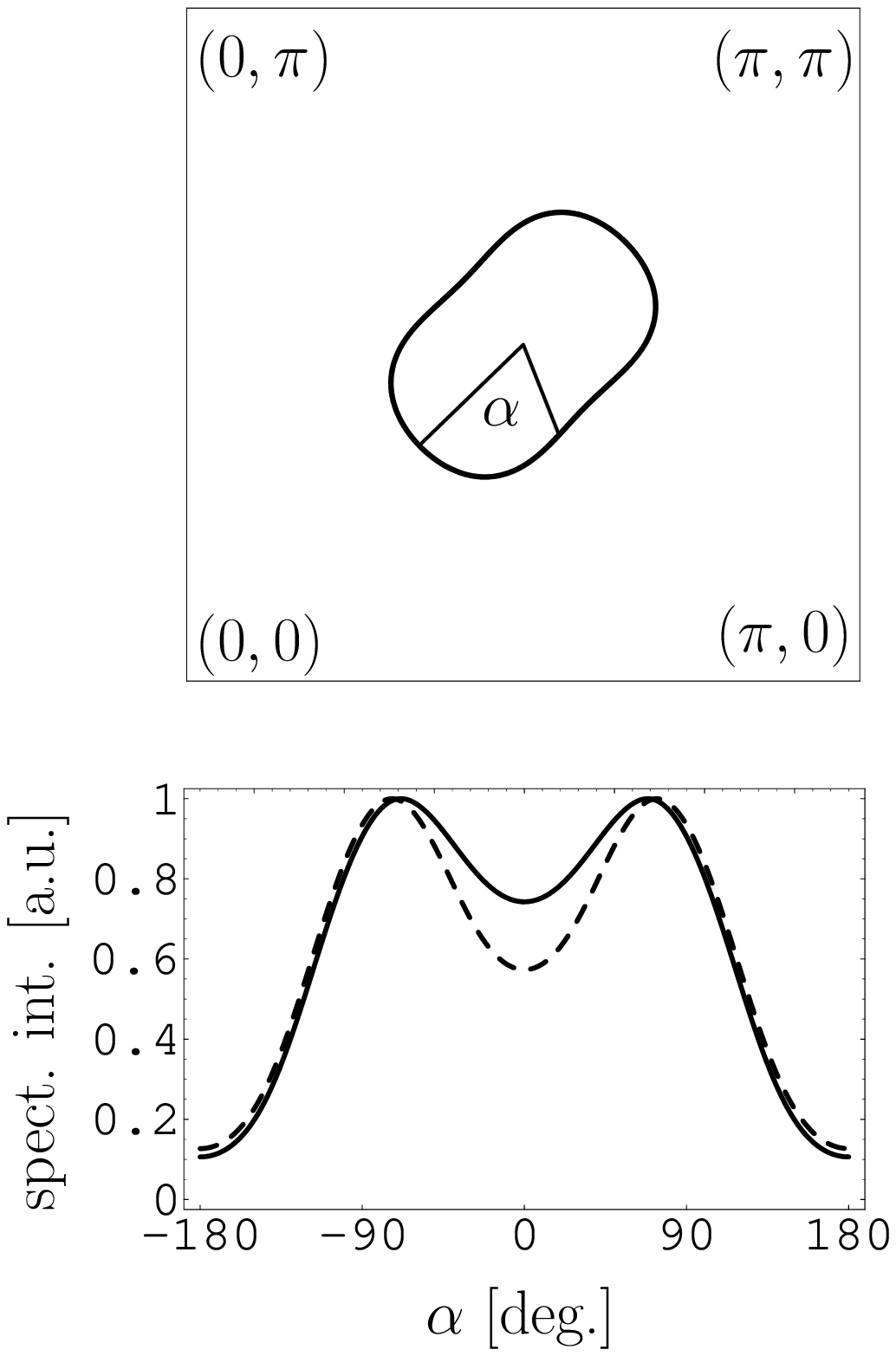}
\caption{\label{fig7} 
Top: Fermi surface obtained for $10\%$ hole concentration
by rigid filling of the lowest SP-band.
Bottom: Spectral weight of the corresponding SP-band
along the Fermi surface as a function of the angle $\alpha$.}
\end{figure}

\begin{figure}
\includegraphics[width=\columnwidth]{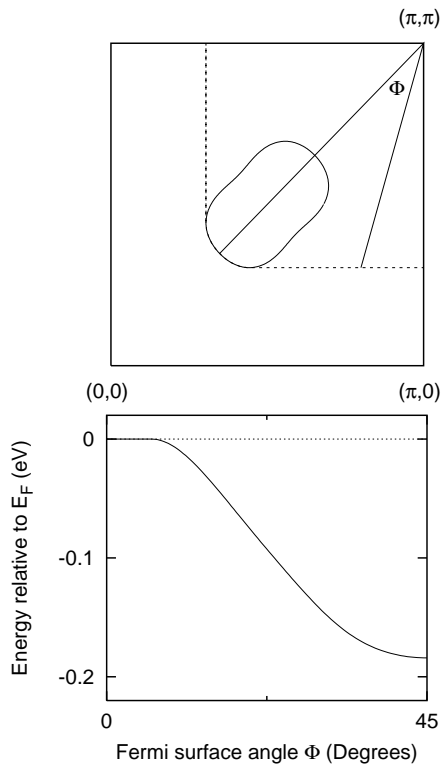}
\caption{\label{fig8} 
Top: Fermi surface obtained for $10\%$ hole concentration
by rigid filling of the lowest SP-band and a contour
obtained by extending the `inner part' of the pocket
to a free-electron-like Fermi surface.
Bottom: Energy relative to Fermi energy of the SP band
forming the pocket
along the free-electron-like Fermi surface as a function of
Fermi surface angle.}
\end{figure}

In a first calculation we want to study the low energy band structure.
In the LCAO-like ansatz (\ref{lcao_like})
we first restrict ourselves to the lowest state (i.e. $m=1$)
for each symmetry $o$ so that we have to solve $4\times 4$ matrices
(we have $o\in \{s,p_x,p_y,d_{x^2-y^2}\}$). 
The resulting band structure is shown in figure \ref{fig5}.
Using the photoemission matrix elements (\ref{pesmat}) the
photoemission spectrum can be calculated, see figure \ref{fig6}.  In
addition to the familiar quasiparticle band 
discovered in Sr$_2$CuO$_2$Cl$_2$ by Wells {\em at al.}\cite{Wellsetal93},
which has been discussed extensively in the 
literature\cite{KyungFerrel,LemaAligia,Chernyshevetal94,Belinicheratal95,Balaetal95,Plakidaetal97,Belinicheretal97,Sushkovetal97,Nazarenkoetal,Leungetal}
there is a second band with slightly lower intensity
which has predominant $p$-like SP character and runs essentially
parallel to the original band. Indeed such a second band with weaker
intensity which follows the main band is seen in exact diagonalization
of small clusters - see e.g. Figure 1 of Ref \cite{EOS}. These higher
lying SP bands have small spectral weight and therefore could be hard
to observe in ARPES - even more so because the ARPES spectra in the
undoped compound are likely to show strong lattice polaronic
effects\cite{polarons}.  Nevertheless, it may be that a
higher lying band - possibly the first $p$-like band -
has been observed by Ronning {\em et al.} in the
insulating cuprate Ca$_2$Cu0$_2$Cl$_2$\cite{Ronningetal05}.
There a second weak band has been observed at roughly
$0.5eV$ below the quasiparticle band. Ronning {\em et al.}
interpreted this as part
of a wide band which reaches $\Gamma$ at a binding energy below $2eV$.
As can be seen in Figure \ref{fig5}, however, this
band (triangles) has a dispersion that is quite
comparable to that of the quasiparticle band (circles).
Moreover, as will be discussed below, 
we believe that the band portion observed in
Ca$_2$Cu0$_2$Cl$_2$ around $\Gamma$ at binding energies
around $2eV$ is an example
of a `1eV peak' as observed in Sr$_2$CuO$_2$Cl$_2$\cite{Pothuizenetal}
and thus unrelated to the band marked by triangles.
While the higher lying bands
may be hard to observe in ARPES optical
interband transitions between the main band and these higher lying
bands produce finite-frequency optical conductivity which may
correspond to the mid IR bands - again in the actual compounds this
may be complicated by polaronic effects.
Another feature which can be seen in Figure
\ref{fig6} and which is quite consistent with experiment is the
sharp drop of the spectral weight of the quasiparticle
band which occurs whenever one passes (roughly) through
the Fermi surface of the noninteracting half-filled. i.e.
the remnant Fermi surface.
It is caused by the ${\bf k}$-dependence
of the photoemission matrix elements (\ref{pesmat}) and
(\ref{pesmat1}) and therefore reflects the interplay of spin and charge
fluctuations in the spin background and the
internal structure of the quasiparticle. 
Assuming that the structure of the quasiparticles
remains roughly the same in a spin background 
without long range order but short ranged antiferromagnetic
correlations - as is suggested by exact diagonalization\cite{bags} -
the ${\bf k}$-dependence
of the photoemission matrix element should be similar
for finite doping. This would provide
an immediate explanation for the `Fermi arcs' seen in the underdoped
compounds.  To make this more quantitative we have
computed the Fermi contour by filling up the single
hole dispersion according to the Pauli principle - see Figure
\ref{fig5} - and showed it in the 
upper panel of Fig. \ref{fig7}. The lower panel 
shows the spectral weight of the quasiparticle band as a function
of the Fermi surface angle. The hole pocket is actually more elongated
along the (1,1) direction than along the
antiferromagnetic zone boundary. It should be noted, however, that this is
calculated at half-filling where $(\pi,0)$ is far from
the valence band top due to the $t'$ and $t''$ terms - see 
Figure \ref{fig5}. On the other hand it is known that upon doping
the band portions near $(\pi,0)$ move upward\cite{Damareview}.
It has been suggested that decreasing antiferromagnetic spin correlations
lead to an effective downward renormalization of
$t'$ and $t''$ with doping\cite{EOS} so that for finite doping
the pocket is probably elongated more along the antiferromagnetic
zone boundary as observed by Chang {\em et al.}\cite{Changetal}.
This effect cannot be reproduced by our simple theory, however.
Another phenomenon which would find a very simple explanation
in an approximate rigid band behaviour upon doping is the
pseudogap\cite{Damareview}. The upper part of
Figure \ref{fig8} shows again the hole pocket and
a contour in ${\bf k}$-space which extends the `front part'
of the pocket into a free-electron-like Fermi surface.
The pseudogap is usually defined by measuring the leading
edge shift or the dispersion of the high-energy feature along such
a contour. The lower part of Figure \ref{fig8} then shows
the energy of the quasiparticle band along this
`Fermi surface' plotted versus the Fermi surface angle $\Phi$.
There is the characteristic flat part near $\phi=0$ which originates because
the contour initially follows the hole pocket and then the
d-wave-like downward dispersion as the free-electron-like Fermi
surface departs from the hole pocket. It should be noted that
this would explain only the high-energy pseudogap.
On the other hand the low-energy pseudogap, being defined
in terms of a leading-edge shift, has no immediate connection
with a dispersion relation and will almost certainly depend also
on the temperature and momentum dependent linewidth of the 
quasiparticle band\cite{Storeyetal2007}.
In fact, assuming a lifetime broadening 
$\Gamma({\bf k}) \propto | E({\bf k}) -\mu|$ would immediately
explain also the low-energy pseudogap.
As noted above the pocket in Figure
\ref{fig8} is too much
elongated in $(1,1)$-direction. Howvere, the effective
downward renormalization of $t'$ and $t''$ in the doped
compounds would lead to a pocket that is
more elongated along the magnetic zone boundary. In any way, however, 
if one would go along the inner part of the pocket
near $(1,1)$ and extend this to a free-electron-like Fermi surface
as in the top part of Figure \ref{fig8}
one will always see a dispersion as shown in the bottom part
of Figure \ref{fig8}.

\begin{figure}
\includegraphics[width=\columnwidth]{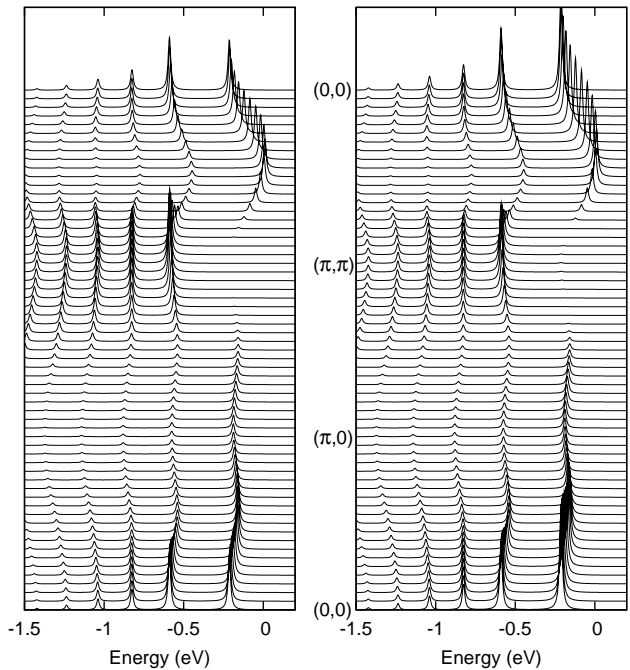}
\caption{\label{fig9} 
Spectral density for hole creation at half-filling
along high-symmetry lines in the Brillouin zone.
Calculated with (right) and without (left) the contribution
from charge fluctuations.}
\end{figure}

Finally we want to discuss the spectral function on a larger energy
scale. To keep the discussion simple we keep only $s$-like SP states
in (\ref{lcao_like})
but include excited $s$-like states with $m=1\dots 10$
(extending this to $m=1\dots 20$ produces no visible change in the
spectra - this is one of the beneficial effects of the
`prediagonalization' of $H_0$). Moreover we
retain only the matrix elements due to string truncation
and the $t'$ and $t''$-terms, i.e. processes of the type
shown in Figure \ref{fig3}. However, we do include
the full photoemission matrix elements (\ref{pesmat}) and
(\ref{pesmat1}). The resulting spectral function
is shown in Figure \ref{fig9}.
In addition to the quasiparticle band - which is
essentially identical to the one of the more exact
calculation shown in Figure \ref{fig6} 
above - there now appear additional
bands at higher energy.  In reality these `bands' probably
are not well-defined states because they are already high in energy.
Rather these states
are probably strongly broadened due to interaction with magnons
and phonons and may have only the character of `resonances'.
Whereas the quasiparticle band is composed mainly of 
the lowest $s$-like SP states for motion of the
hole trapped in the linearly ascending potential the higher lying
bands correspond to excited levels of the trapped hole.
The relatively high intensity of these states may be understood by
noting that the coefficients $\alpha^{(m)}_\nu$ for the excited
states $m>1$ will have extra nodes as functions of $\nu$ and if the 
signs of the
$\alpha^{(m)}_\nu$ better match the prefactors in
(\ref{pesmat}) the matrix element may even be larger for
these higher lying states.\\
The spectral function in Figure \ref{fig9} is qualitatively similar
to the result of a recent calculation by
Bonca {\em et  al.}\cite{Boncaetal07} which was performed in a string
basis with several million basis states. As already
noted the main difference
between the present calculation and the one by Bonca {\em et  al.}
is in the fact that the `prediagonalization'
of $H_0$ by solution of (\ref{schroedinger}) leads to
a quite massive reduction of irrelevant degrees of freedom
in the present scheme - as noted above, the matrix diagonalized here
is $10 \times 10$. Despite this simplification
not only the dispersion of the topmost peak but also the fact that the
high energy part `widens' as one moves away from $(0,0)$
is reproduced. On the other hand, it should be also noted that the 
spectral weight
shown in Figure \ref{fig9} cannot directly be compared with
Ref. \cite{Boncaetal07} because we are taking into account the
corrections of the  ARPES matrix element
due to quantum spin fluctuations in the half-filled gound state
which are not included there.

\section{Electronic structure at higher binding energies}

Recently a number of studies have revealed additional
structure in the ARPES spectra at higher binding energies.
Ronning {\em at al.} found\cite{Ronningetal05} that
in the insulating cuprate Ca$_2$Cu0$_2$Cl$_2$
the quasiparticle band seem to `fade away' as
the $\Gamma$-point is approached - a phenomenon which is common
to all cuprates. The spectral weight that is missing
from the quasiparticle band then appears at $\approx 1.5eV$ below
the maximum of the quasiparticle band in the form a
high intensity band which has a dispersion that is remarkably
consistent with LDA band structure calculations.
More precisely this is the antibonding band of
$Cu3d_{x^2-y^2}$ and $O2p\sigma$ orbitals - whereby 
it has to be kept in mind that at $\Gamma$ these two
orbitals do not hybridize due to partity,
so that in an LCAO-like description these bands have
pure oxygen character at $\Gamma$ and - by continuity in ${\bf k}$ -
very small $Cu3d$ admixture its neighborhood. 

Similar behaviour was observed in doped cuprates as
well\cite{Grafetal07,Xieetal,Vallaetal,Panetal,Changetal07}. 
In addition, the so-called waterfall-phenomenon is observed.
Moving e.g. along the $(1,1)$ direction towards $\Gamma$
the quasiparticle band
first disperses away from the Fermi edge but then -
at a momentum of approximately $(\frac{\pi}{4},\frac{\pi}{4})$ -
seems to bend down sharply and drop almost vertically
down to an energy of $\approx 1eV$ below the Fermi edge.
The apparent vertical part of the dispersion - the `waterfalls' -
can be seen as a hump in momentum distribution curves
at  binding energies in the range of $0.5eV \rightarrow 1.0eV$
Upon reaching $\approx 1eV$ below $E_F$ the vertical parts then 
merge with two LDA-like bands of high intensity.

Similar behaviour - namely small band portions with a free
electron-like dispersion and high spectral weight near
high symmetry points of the Brillouin zone
- has been observed previously in the insulating compound
Sr$_2$CuO$_2$Cl$_2$ by Pothuizen {\em at al.}\cite{Pothuizenetal}.
The interpretation given by these authors was that these are
O2p derived states which do not hybridize with Cu3d orbitals due to
symmetry - which is why they appear only at high-symmetry
points of the Brillouin zone - and hence are unaffected by
the strong correlations in the partially filled
Cu3d orbitals. This explains their
LDA-like dispersion and high spectral weight because they are
essentially free electron states. 

We believe that an important clue to the interpretation of the 
waterfalls is the finding of Inosov {\em et al.} who
showed that matrix element effects play a crucial role
in their observation\cite{Kordyuketal07} and that the
missing part of the quasiparticle band near $\Gamma$ can in fact be
observed with photon energies around $100eV$ where the cross section
for $Cu3d$ orbitals becomes appreciable\cite{Eastman_Freeouf}. 
Moreover, Pan {\em at al.} found that the
waterfalls show the same dependence on photon polarization as
the quasiparticle band itself, indicating that they also are
derived from Zhang-Rice singlets\cite{Panetal}. And finally in our
opinion the crucial clue is the fact, that the quasiparticle band itself
cannot be observed near the $\Gamma$-point either. While numerical
studies of the Hubbard and t-J model\cite{Ortolani,EskesEder96,Groberetal00} 
- as well as the present theory - do indeed predict that the spectral
weight of this band is lower by a factor of 2-3 at $\Gamma$ as compared
to $(\frac{\pi}{2},\frac{\pi}{2})$, in experiment there is
practically no more intensity visible. Rather, the spectra show
a complete suppression of spectral weight around $\Gamma$ which
extends down to the intense LDA-like bands.
Most significantly, however,
the waterfalls appear at very nearly the same momentum where the
quasiparticle band itself becomes visible.

We conclude that the reason for the vanishing of the quasiparticle
band near $\Gamma$ is `extrinsic' to the t-J or single-band Hubbard model 
namely the special combination of phases for the $O2p\sigma$ orbitals
in the bonding combination which hybridizes with a given
$Cu3d_{x^2-y^2}$ orbital in the Zhang-Rice singlet (ZRS).
In the framework of a simple
three-step model of photoemission for photoemission
the effect that we seek
comes from the matrix element for the dipole transition
from O2p states into
the final state - which we take to be a plane wave with momentum 
${\bf k}$ for simplicity. For definiteness
we introduce the dipole matrix element 
\begin{equation}
{\bf \epsilon} \cdot {\bf v}_\alpha = \frac{1}{\sqrt{V}}\;
\int d{\bf r}\; e^{-i{\bf k}\cdot {\bf r}} \;
{\bf \epsilon}\cdot{\bf r}\;
\psi_\alpha({\bf r})
\label{dipole}
\end{equation}
where ${\bf \epsilon}$ denotes the polarization vector of the
light, $V$ the volume of the
crystal, $\psi_\alpha({\bf r})$ is
the wave function of an O2p orbital at the origin and 
$\alpha \in \{x,y\}$. 
The matrix element for a dipole transition from the bonding combination
of O2p$\sigma$ oxygen orbitals around a
given Cu site $j$ 
\begin{equation}
P_{j,\sigma} = \frac{1}{2}(
p_{j+\frac{\hat{x}}{2},\sigma} - p_{j-\frac{\hat{x}}{2},\sigma}
- p_{j+\frac{\hat{y}}{2},\sigma} + p_{j-\frac{\hat{y}}{2},\sigma})
\label{zrphases}
\end{equation}
into the plane wave state is
\begin{equation}
m_{ZRS}=e^{-i{\bf k}\cdot {R}_j}\;
i\left[
{\bf v}_y \;\sin \left(\frac{k_y}{2}\right) -
{\bf v}_x \;\sin \left(\frac{k_x}{2}\right) \right] \cdot {\bf \epsilon}.
\label{Zhangdip}
\end{equation}
Note that the ${\bf k}$ dependence of the expression in square
brackets comes solely from the interplay between the phase factors 
$e^{i{\bf k}\cdot{\bf r}}$ on
the four oxygen neighbors of atom $j$ and the relative phases of the 
orbitals in (\ref{zrphases}).
Namely any two oxygen orbitals whose position in real space
differ by one lattice spacing have a relative phase of $(-1)$
in the ZRS - which would correspond to momentum $(\pi,\pi)$.
This ${\bf k}$ dependence is therefore
completely independent of details in the computation of the
matrix elements ${\bf v}_\alpha$ and therefore in particular
independent of the photon polarization.
Moreover it would stay the same if a more
realistic final state wave function were used as long as this
is a Bloch state with momentum ${\bf k}$.
It now can be seen that $m_{ZRS}\rightarrow 0$ as ${\bf k}\rightarrow
(2n\pi,2m\pi)$. Unlike the argument based on
the dipole selection rule which was discussed by 
Ronning {\em et al.}\cite{Ronningetal05}
the expression (\ref{Zhangdip}) thus explains why the 
quasiparticle band has vanishing intensity not only
at $\Gamma$ but at all equivalent points in higher Brillouin zones
as well.

The above considerations apply only to hole creation on oxygen.
For photon energies around $20eV$ is is well known, however, that
the cross section for hole creation on oxygen is considerably larger
than for hole creation in transition metal $3d$ 
orbitals\cite{Eastman_Freeouf}.
Moreover, starting from a $d^9$ state one would reach
a $d^8$ state. The latter is high in energy
and hence has small weight in the ZRS and moreover
the resulting spectral weight would be spread over
several eV due to the multiplet splitting of $d^8$ so that
the corresponding matrix element certainly is small.
A $d^9$ final state could be produced by photoemission from
$d^{10}$ state but this state also is high in energy
and therefore has small weight in the ground state.
In any way the contribution of hole creation on Cu
is strongly suppressed. For larger photon energies - around $100eV$ - 
Inosov {\em at al.}
could indeed resolve the quasiparticle band\cite{Kordyuketal07} but only
at $(2\pi,0)$.
Obviously the larger photoemission cross section and the
non-validity of the dipole selection rule at this
momentum make it possible to see intensity from photo holes
in $Cu3d_{x^2-y^2}$ orbitals. Clear evidence that the intensity
seen there is due to photohole creation on Cu is also
provided by the strong variation of intensity around the
$Cu3p\rightarrow Cu3d$ threshold which is being used
routinely to identify transition metal $3d$ states in other
transition metal oxides\cite{Ohetal}.

One might then ask, where the spectral weight corresponding to hole
creation on oxygen goes at $\Gamma$. The answer is that
this spectral weight is concentrated in the
LDA-like bands which are observed around the $\Gamma$-point
at approximately $1.5 eV$ below the Fermi energy.
Due to parity the `antibonding band' of
$Cu3d_{x^2-y^2}$ and $O2p\sigma$ orbitals actually has
pure oxygen character at $\Gamma$. If a Bloch state of $O2p$ orbitals is
created with momentum $(0,0)$ this cannot couple to the ZRS
singlet - see (\ref{Zhangdip}) - but has overlap $1$
with these bandlike states. This is the same reasoning as given 
by Pothuizen {\em et al.}\cite{Pothuizenetal} for the `1eV peaks' 
in Sr$_2$CuO$_2$Cl$_2$. The spectral weight transfer from the
quasiparticle band to the `1eV peaks' at $\Gamma$ as observed
by Ronning {\em et al.} thus is strong evidence for the ZRS-character
of the quasiparticle band.
\begin{figure}
\includegraphics[width=\columnwidth]{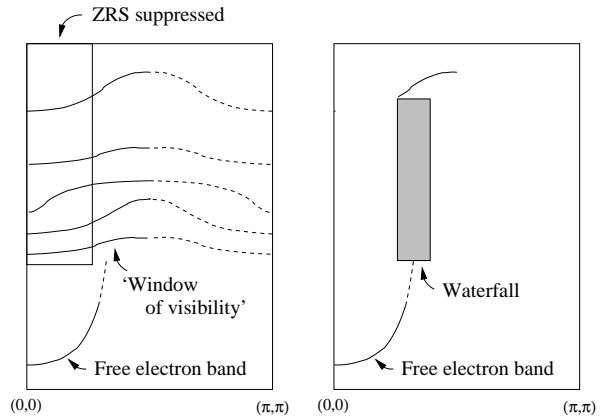}
\caption{\label{fig10} 
Interpretation of the waterfall phenomenon.
Around $\Gamma$ photoholes on oxygen do not couple
to the ZRS whence all t-J derived states
have no spectral weight. Photoholes created
with this momentum on oxygen instead propagate as (nearly)
free-electron
states which do not hybridize with the correlated
Cu d-orbitals.
Moving away from $\Gamma$ the t-J bands become visible
but lose spectral weight as well (dashed potions).
The nearly free-electron states cease to exist because
they now have appreciable hybridization with the correlated
Cu d-orbitals. The higher lying SP bands thus are
visible only in a small window in ${\bf k}$-space i.e.
the waterfalls.}
\end{figure}

These leaves the question as to what is the interpretation of the
waterfalls. The above considerations suggest
that these are simply the higher lying SP bands
seen in Figure \ref{fig9}. These bands
are likely to be strongly broadened due to interaction
with spin excitations and possible also phonons although
one of them may have been resolved by 
Ronning {\em et al.}\cite{Ronningetal05} (see Figure \ref{fig5}).
Near $\Gamma$ the spectral weight of these bands disappears
because they are also `t-J-derived' and the
matrix element for creation of a ZRS 
vanishes see Figure \ref{fig10}. As one moves away from $\Gamma$ the matrix
element for creation of a ZRS increases
but - as can be seen in Figure \ref{fig9} - the spectral
weight of these excited SP bands now quickly decreases.
It follows that the spectral weight of these bands must
go through a maximum as one moves away from $\Gamma$ and this is
our interpretation of the `waterfalls': a number of essentially
incoherent states which have a `window of visibility'
in a narrow range of momenta around $(\frac{\pi}{4},\frac{\pi}{4})$
and this window of visibility is seen as the hump in the
momentum distribution curves. Since the suppression of these states
near $\Gamma$ is governed by the same matrix  element of
the ZRS as the quasiparticle band itself, it is
moreover clear that this window of visibility `opens' precisely
in the same range of ${\bf k}$
where the quasiparticle band itself becomes visible - hence the
apparent downward bending and the waterfall-like appearance of 
the spectra. 

A more quantitative description of this phenomenon
obviously would have to start out from a three-band
model so as to describe both, the coupling
of a photohole to a ZRS and the existence of the nearly free electron
states at high-symmetry points. This is out of the scope
of the present paper and we therefore make no attempt for
a quantitative discussion.
There have been a number of attempts to explain the waterfall
phenomenon within the t-J model\cite{Manousakis07,Zemljicetal07}. 
However, as was already noted, the experimental spectra show a 
complete suppression of spectral weight around $\Gamma$ which extends
downward all the way to the intense LDA-like bands. This behaviour is not
reproduced by t-J model calculations which 
show spectral weight - corresponding to the 
higher lying bands in Figure \ref{fig9} -
at too low binding energy. Moreover, the LDA-like dispersion
of the high-intensity parts near $\Gamma$
is not really reproduced

\section{Optical conductivity}
\begin{figure}
\includegraphics[width=\columnwidth]{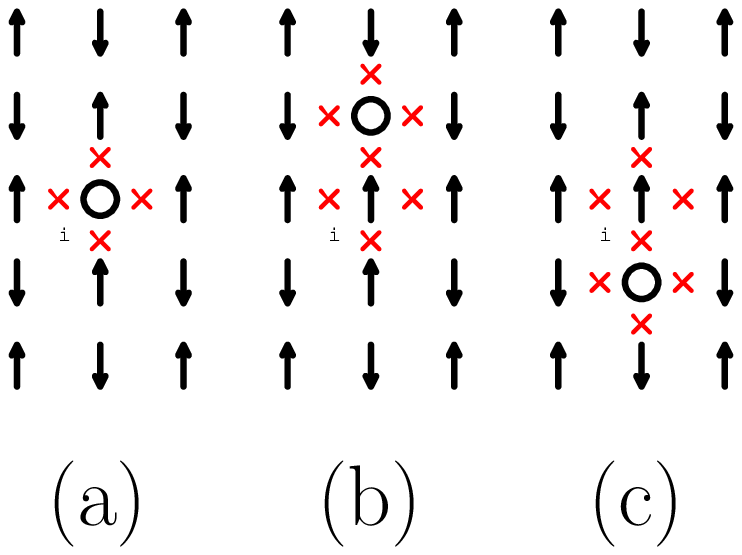}
\caption{\label{fig11} 
Application of the current
operator $j_y$ to  a bare hole in the N\'eel state (a)
creates two strings of length $1$ (b) and (c)
with opposite signs.}
\end{figure}

\begin{figure}
\includegraphics[width=\columnwidth]{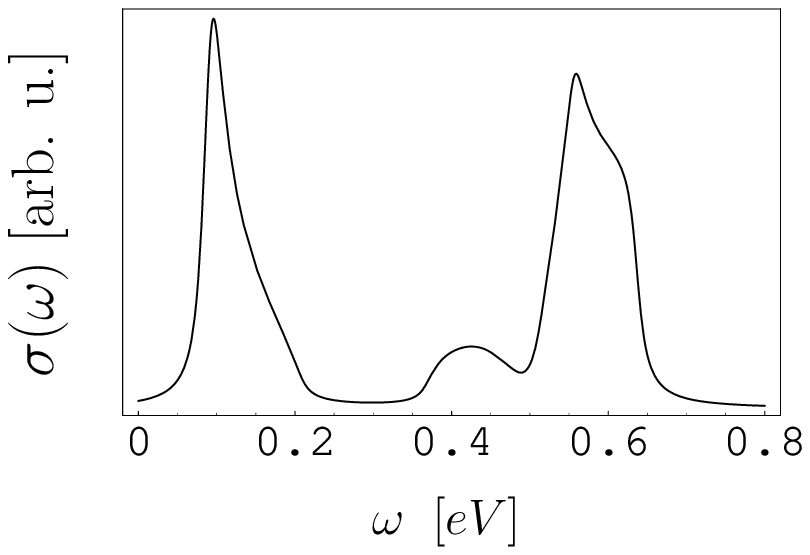}
\caption{\label{fig12} 
Optical conductivity at 10 \% hole concentration.}
\end{figure}

Finally we turn to a discussion of the optical conductivity
thereby using the results of the simplified calculation
which took into account
only the lowest $(m=1)$ state for each symmetry $o$.
As is the case for atomic wave functions the
$s$ and $d$-like SP states on one hand and the
$p$-like SP states on the other have opposite parity
and hence can have nonvanishing matrix elements of the
current operator in between them. If we assume again that the
quasiparticle band is filled with holes upon doping  - which would 
occupy momenta around $(\frac{\pi}{2},\frac{\pi}{2})$ -
we thus
expect optical interband transitions (as indicated in Figure
\ref{fig5}) which should be observable
in the finite frequency optical response. This is defined as
\begin{eqnarray*}
\sigma_\delta ({\bf q}=0,\omega)&=&
\sum_{n\neq 0} \frac{1}{\omega}\;|\langle\psi_n|j_\delta({\bf q}=0)
|\psi_0\rangle|^2\\
&& \;\;\;\;\;\;\;\;
\; \delta( \omega - (E_n - E_0) ).
\label{opcon}
\end{eqnarray*}
where $|\psi_n\rangle$  ($E_n$ ) denotes the n-th eigenstate (eigenenergy) of
the system (in particular, $n=0$ denotes the ground state).
Also, $j_\delta$, with $\delta=x,y$ denotes a component of the current
operator,
\begin{equation}
{\bf j}({\bf q}) = i
\sum_{m,n} t_{mn}
e^{i {\bf q}\cdot ({\bf R}_m + {\bf R}_n)/2 }
\;[{\bf R}_m - {\bf R}_n\;]
\hat{c}_{m,\sigma}^\dagger \hat{c}_{n,\sigma}.
\end{equation}
Assuming a filling of the quasiparticle band we approximate this by
\begin{eqnarray}
\sigma_\delta (\omega) &=&
2 \sum_{{\bf k}} \sum_{l=1}^3 \frac{1}{\omega}\;
|\langle \Phi_l({\bf k})|j_\delta|
         \Phi_1({\bf k})\rangle|^2 n_{{\bf k}}\nonumber \\
&& \delta( \omega - ( E_l({\bf k}) - E_1({\bf k}))).
\end{eqnarray}
where $n_{\bf k}$ denotes the hole occupation of the
quasiparticle ($l=1$) band. To evaluate this we need matrix elements
of the current operator between the localized
SP states (\ref{wvfn}): 
$\langle \Psi^{\lambda}_i | j_\delta | \Psi^{\lambda'}_i\rangle$
(with $\lambda=(o,m)$).
The main contribution to this matrix element are shown in
Figure \ref{fig11}. Starting from a bare hole - the `string of length 0' -
the current operator $j_y$ generates two strings of length $1$
but with opposite sign. These two string states thus have precisely
the right sign to couple to a $p_y$-like SP state, see the
sign convention in Figure \ref{fig2}. From this and similar processes
we obtain the matrix element
\begin{equation}
\langle \Psi^{(p_{\delta},0)}_i | j_\delta |\Psi^{(s),0}_i
\rangle=
2 i  (\sum_{\mu=0}\alpha_\mu
\alpha^{\prime}_{\mu+1}-\sum_{\mu=1}\alpha^{\prime}_{\mu}\alpha_{\mu+1}).
\end{equation}
All matrix elements generated in a similar way and involving SPs of different symmetries or hopping
at longer distances have been listed in the Appendix \ref{appb}.
The resulting optical conductivity at a hole concentration of 10 \%
is shown in Figure \ref{fig12}.
While the higher lying bands with $p$-character have small spectral
weight and hence are difficult to observe in ARPES, they dominate the
optical conductivity.  We note that the present interpretation of the
optical conductivity is consistent with that of Ref. \cite{Ederetal96}
and \cite{VojtaBecker97}.  One may expect that the excited levels of
the trapped hole - or `internal degrees of freedom' of the SP
quasiparticles - are also seen in the dynamical
density correlation function.
Vojta and Becker\cite{VojtaBecker97} have calculated the
dynamical density correlation function in
a string framework similar to the one used by Bonce {\em at al.} and
obtained convincing agreement with exact diagonalization results.
Within our approach one can naturally explain the distribution and
spreading of the spectral intensity over a wide energy range, which
has been observed in experiments \cite{Uchidaetal91,Hwangetal07}. Due
to the Brillouin folding in the AF state the center of the hole pocket
at $(\pi/2,\pi/2)$ is a high symmetry point. $s$, $d$, and $p$ SP states
do not mix with each other exactly at this point. Thus, the transitions
with similar intensity occur in the weakly doped system between the
lowest predominantly $s$-like and the first and the third 
predominantly $p$-like
bands. Wider spreading of the high intensity region at higher energies
can be attributed to the admixture of $p$-wave SPs to the lowest band
and transition to the predominantly $d$-wave second excited band.

\section{Discussion and conclusions}

In summary we have presented a theory for spin polaron-like
quasiparticles. The basic idea is that a hole in an
antiferromagnetically ordered `spin background' is self-trapped
which actually requires only short-range antiferromagnetic
order. This will lead to a hierarchy of localized states
which may also realize different irreducible representations of
$C_{4v}$. Since the Heisenberg exchange and the
$t'$ and $t''$ terms have matrix elements between such
self-trapped states on neighboring sites an LCAO-like
description emerges where the role of the atomic or
Wannier functions is played by the levels of the self-trapped
hole. This leads to a multi band structure for the doped holes
with the lowest of these being the familiar 
quasiparticle band observed in the insulating compounds
and discussed extensively in the literature.
Here we take the point of view that the simplest description
for the underdoped compounds is holes being filled into this
quasiparticle band.
The fact that these self-trapped states extend over several
unit cells in real space necessarily implies that they have
an ARPES form factor which varies within the first Brillouin zone - 
hence the strong variation of the photoemission intensity
of the quasiparticle band as a function of ${\bf k}$
which explains the remnant Fermi surface and the Fermi arcs seen 
in ARPES. Moreover, the pseudogap becomes a triviality
within this picture.
One of the higher bands of the effective LCAO-Hamiltonian
may have been observed in ARPES in the insulator Ca$_2$CuO$_2$Cl$_2$
and optical transitions between the resulting bands may
explain the mid-infrared band in optical spectroscopy.
There are probably complications due to lattice polaron effects
but here we neglect these although the simplicity of the
present calculation - which never need matrices to more than
$10\times 10$ - would certainly allow to treat such effects as well.

The main drawback of the present theory is the use
of a spin background with antiferromagnetic order - which clearly
is not realized in doped materials of interest. On the other hand
all processes by which the hole propagates involve only spins in its
immediate neighborhood. One may therefore expect that very similar
processes would occur in a spin background with only short range
antiferromagnetic correlations so that much of the present theory
should apply in this case as well. The most important effect
we are missing with the present calculation is the closing of the pseudogap
with both increasing temperature and increasing doping.
Since this closing implies that the dispersion actually approaches
that for the `pure' t-J model - i.e. without $t'$ and $t''$ terms -
this effect could be described by an effective
downward renormalization of the $t'$ and $t''$ terms.
The mechanism may be the decrease of spin correlations: in the
N\'eel state the $t'$ and $t''$ terms can transport a hole 
`completely coherent' i.e. without creating a spin excitation. 
As the spin correlation length decreases and reaches the `range' of 
these terms in real space - two lattice spacings -
the $t'$ and $t''$ will increasingly generate spin excitations
when transporting a hole so that they change their net effect
from coherent hole transport to `excitation generating' hole
transport.

\begin{acknowledgments}
P.W. acknowledges encouragement and valuable comments from P. Fulde and R. Micnas.
He also appreciates discussions with M. Jarrel, P. Prelov\v{s}ek, and T. Tohyama.
\end{acknowledgments}

\begin{appendix}
\section{Spin polaron model}
\label{appa}
We first derive the equations for the
coefficients $\alpha^{(o,m)}(l)$ in (\ref{wvfn}), thereby assuming
(\ref{prefs}).
We denote by $|\Phi_\nu\rangle$ the sum of all string states
with $\nu$ inverted spins, each multiplied by the proper phase factors
according to figure \ref{fig2}.
On a Bethe lattice the number of such states is
$n_\nu=f^{(o)}\;z(z-1)^{\nu-1}$ where $f^{(o)}=1$ for 
$o=s,d_{x^2-y^2}$ and $f^{(o)}=1/2$ for $o=p_x,p_y$.
Our normalized basis states are 
$|\nu\rangle = n_\nu^{-1/2}|\Phi_\nu\rangle$
which obey
\[
\langle \nu+1|H_t|\nu \rangle = t\; \sqrt{\frac{n_{\nu+1}}{n_\nu}}
\]
because each of the $n_{\nu+1}$ basis string states in the bra is
generated exactly once from a state in the ket and the matrix element
for hopping of a hole is positive. The states (\ref{wvfn}) can be written as
\[
|\Psi_i\rangle = \sum_\nu \beta_\nu |\nu\rangle
\]
with $\beta_\nu= n_\nu^{1/2} \alpha^{(o,m)}_{\nu}$.
Performing the variational procedure - thereby using (\ref{ising}) -
we obtain the set of equations
\begin{eqnarray}
E_0 \beta_0 + \sqrt{z}t \beta_1 &=& E\beta_0 \label{schr1}\\
E_1 \beta_1 + \sqrt{z}t \beta_0 + \sqrt{z-1}t \beta_2 &=&
E\beta_1 \label{schr2}\\
E_\nu \beta_\nu + \sqrt{z-1}t (\beta_{\nu-1} + \beta_{\nu+1}) &=&
E\beta_\nu \label{schr3}
\end{eqnarray}
For $o=p_x,p_y,d_{x^2-y^2}$ one has to set $\beta_0=0$ and discard
the first equation. After introducing a cutoff for
$\nu$ these equations can be solved numerically.

Now we proceed to analyse with some sample processes which give rise
to nonvanishing matrix elements (\ref{hmatel}) and (\ref{ovmatel}).
The hopping integral $t$ is the highest energy scale among all model
parameters $t$, $t'$, $t''$, and $J$. On the other hand, the exchange
energy grows fast with the number of fluctuations. That number is
directly related to the length of paths ${\cal P}_i$.  This mechanism
is responsible for the tendency towards hole confinement and the
construction of SPs also relies on it.  Quasiparticle deconfinement
occurs due to processes which are mediated by hopping to second and
third NN and by the action of the XY term in the exchange interaction.
Only string states $|{\cal P}_i\rangle$ with a low number fluctuations
can be involved in those processes at the low energy scale.  Thus, in
order to find matrix elements (\ref{hmatel}) we need to determine
coupling between short-string states $|{\cal P}_i\rangle$. That
coupling is induced by the perturbation part $H_1$ which by definition
contains terms related to hopping to second and third NN sites and
which also contains the XY term. We restrict our considerations to
processes involving
string states $|{\cal P}_i\rangle$ which are related to paths not
longer than 2 lattice spacings for $s$-wave SPs and to paths not
longer than 3 lattice spacings for SPs with lower symmetry. In the
latter case we take into account longer strings because the
zero-length string state $|{\cal P}_i\rangle$ does not contribute to
the wavefunction (\ref{wvfn}) of SPs with lower symmetry and thus the
weight is shifted to states representing longer paths.  The absolute
values of prefactors $\alpha_{\mu}$, $\alpha^{\prime}_{\mu}$
corresponding to lowest eigenstates of (\ref{schr1})-(\ref{schr3})
rapidly decrease with the string length $\mu$ which additionally
justifies the restriction of our considerations to strings with short
length. At the level of the approximation which we apply, there exist
more than 20 categories of processes which contribute to
(\ref{hmatel}) and (\ref{ovmatel}). The differences between different
categories concern the underlying mechanisms or the geometry of
involved strings. Since the mechanisms which give rise to coupling
between $s$-wave SPs were discussed in detail in the past
\cite{EderBecker90,WrobelEder98} here we will mainly concentrate on
the issue how the lowering of SP symmetry influences the coupling
between SPs and we will discuss representative examples of processes
which give rise to the hybridization between SPs. In order to keep the
Hilbert space as small as possible we consider at first the low energy
sector and analyze only lowest SP states with given symmetry.

\begin{figure}
\includegraphics[width=\columnwidth]{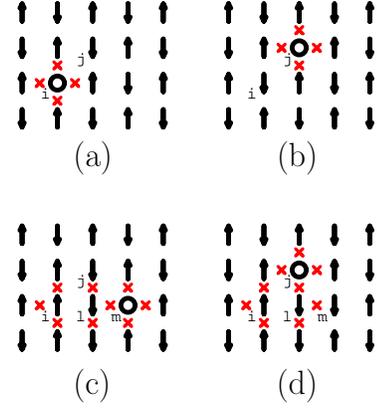}
\caption{\label{fig13} 
(Color online) (a) and (b) A process which gives rise to the
shift of the $s$-wave SP to a second NN site. (c) and (d) A process
which generates corrections to eigenenergies $E_1$ and $E^{\prime}_1$
of SPs.}
\end{figure}

For example, Fig. \ref{fig13} (a), (b) depicts a process during which a SP
polaron is shifted to a second NN site. 
(a) represents a string state $|{\cal P}_i\rangle$ which is a
component of the wavefunction (\ref{wvfn}) of a SP created at site
$i$. This state has been obtained by creating a bare hole in the
N\'eel state. It appears with the prefactor $\alpha_0$ in the
superposition (\ref{wvfn}) defining the $s$-wave SP at site $i$.  With
the same prefactor appears the string state depicted in
Fig. \ref{fig13} (b). It is a component of the $s$-wave SP created
at site $j$. Since string states (a) and (b) are coupled by the second NN
hopping term a non-vanishing matrix element (\ref{hmatel}) between
$s$-wave SPs is generated. The related contribution to that matrix
element is given by
\begin{equation}
\delta{\cal T}^{((s,0)(s,0))}_{ji}\equiv\delta{\cal
  T}^{(ss)}_{{\bf R}_j-{\bf R}_i}=\tau^{(1)}_{ss}=t^{\prime} \alpha_0^2.
\label{tau1}
\end{equation}
Within a convention for presenting matrix elements this contribution can
be written as 
\begin{equation}
ss,\hat{\bf x} +\hat{\bf y}, \tau^{(1)}_{ss}:\;C_4.
\label{con1}
\end{equation}
 The
presence of the group symbol $C_4$ in (\ref{con1}) means that
additional contributions to (\ref{hmatel})
 can obtained by applying to (\ref{tau1}) elements of $C_4$
different than identity.   Since $\alpha^{\prime}_{0}\equiv 0$ states
depicted in Fig. \ref{fig13} (a) and (b) do not contribute to SP
wavefunctions (\ref{wvfn}) with lower symmetry. Thus the above
discussed lowest-order process does not generate hybridization in
which such SPs are involved. By process order we understand, in terms
of the conventional perturbation theory, the total number of
Hamiltonian actions necessary to transform a state representing
a bare hole created in the AF background into another such state. It is
clear that this parameter is related to the sum of lengths for paths
involved in a given process.

The origin of the formula for the integral $\tau^{(1)}_{ss}$ in
(\ref{tau1}) is rather obvious. It is the product of prefactors which
appear by string states depicted in Fig. \ref{fig13} (a) and (b) in
the definition for the wavefunctions of SPs at sites $i$ and $j$. $t'$
is the integral which appears as a prefactor in the hopping term
coupling  those string states. The same scheme can be applied to deduce
the form of hopping integrals which appear in further terms in
${\cal
  T}^{(oo')}_{{\bf R}}$.

\begin{figure}
\includegraphics[width=\columnwidth]{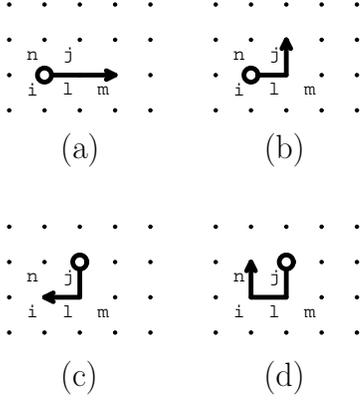}
\caption{\label{fig14} 
(a) and (b) A different representation of 
string states depicted in Fig. \ref{fig13} (c) and (d)
respectively. (c) A string state contributing to a process which
involves hopping to second NN sites. That process gives rise to SP
shifts along plaquette diagonals. (d) A string state contributing to a process which
involves hopping to NN sites.}
\end{figure}

Fig. \ref{fig13} (c) and (d) depicts a process which gives rise to
corrections to SP eigenenergies $E_1$ and $E^{\prime}_1$ of SPs and
amendments to the diagonal term in (\ref{hmatelksp}). Those corrections
originate with the coupling between different string states generated
by the hopping to second NN sites. Those string states contribute to
the wavefunctions of SPs created at the site $i$. The paths ${\cal
P}_i$ corresponding to string states ${\cal P}_i$  depicted
in Fig. \ref{fig13} (b), (c) have been shown in
Fig. \ref{fig14} (a), (b), respectively. From now on we will also
use the latter form of diagrams to represent string states.
The same mechanism as discussed above works for pairs of longer
strings which have identical forms with the exception of hole
positions at ends of them. The corrections to energies of $s$-wave
$d$-wave and $p$-wave SPs are $\tau^{(13)}_{ss}$, $\tau^{(13)}_{dd}$
and $\tau^{(13)}_{pp}$ respectively. Their explicit form can be found
below. Due to point-group properties the above
discussed process does not generate the coupling between SPs with
different symmetries. 

During another process which also involves strings of length 2
lattice spacings the action of hopping to second NN sites 
generates shifts of SPs between second NN
sites. Surprisingly, SPs are shifted now in the direction opposite to the
direction of the bare hole move. The nature of that paradox can be
understood by analyzing Figures \ref{fig14} (b) and (c). The action
of the second NN hoping term on the string state, Fig. \ref{fig14}
(b), which contributes to the SP created at site $i$, shifts the hole
at the end of that string from site $j$ to site $i$ and generates a
string state, Fig. \ref{fig14} (c), which contributes to the SP
created at site $j$.  That process gives rise to a new contribution to
${\cal
  T}^{(oo')}_{\bf R}$,
\begin{eqnarray}
&&ss,\hat{\bf x}+ \hat{\bf y},
\tau^{(15)}_{ss};\;dd, \hat{\bf x}+ \hat{\bf y},
\tau^{(15)}_{dd};\; \nonumber\\&& p_yp_x,\hat{\bf x}+ \hat{\bf y}, \tau^{(15)}_{pp}: \; C_{4v} 
\label{del2ahef} \\
&&
p_ys,\hat{\bf x}+ \hat{\bf y},\tau^{(15)}_{ps};\;
p_yd, \hat{\bf x}+ \hat{\bf y}, \tau^{(15)}_{pd}: \;C_{4v} \cdot H.C.
\label{del2hef}
\end{eqnarray}
The presence of the group symbol $C_{4v}$ means as
before that the contributions (\ref{del2ahef}) and (\ref{del2hef}) should be supplemented
by matrix elements obtained by applying to the existing elements all group
elements and  in the case of the second line also elements
obtained by applying the Hermitian conjugation. The process
depicted in Fig. \ref{fig14} (b) and (c) gives rise to the
hybridization of SPs with different symmetries because apart from the case
of coupling between $s$-wave and $d$-wave SPs no
selection rule forbids it and also because string states
depicted in both Figures \ref{fig14} (b) and (c) can be components of
wavefunctions for SPs of arbitrary symmetry. The hopping integrals
$\tau^{(15)}_{oo'}$ as usually are given by products of the bare
hopping integral $t'$ and of prefactors with which string states
depicted in Fig. \ref{fig14} (b) and (c) appear in the definition of SPs. The
existence of negative signs in some of integrals $\tau^{(15)}_{oo'}$
originates with the fact that due to symmetry properties of some SPs
such signs are also contained in prefactors of string states in sums
(\ref{wvfn}) defining those SPs.

After discussing some representative 	examples of SP coupling generated
by hopping to second NN sites we skip the discussion of analogous processes
mediated by hopping to third NN sites, because their mechanisms are
similar, and analyze a high-order process which involves hopping to NN
sites. That process has been neglected during the construction SPs,
because a long-string state participates in it. The NN hopping
operator, by shifting the hole at the end of the string depicted in
Fig. \ref{fig14} (d) from site $n$ to site $j$, generates the string
state depicted in Fig. \ref{fig14} (b), which gives rise to a new
contribution to the coupling between SPs created at sites $i$ and
$j$. For the sake of clarity, we only additionally mention briefly, that
Fig. \ref{fig14} (d) represents a string state in which flipped
spins occupy sites $j$, $l$, $i$, and the hole occupy site $n$.  It is
clear that the same mechanism generates the hopping of SPs between all
second NN sites. Depending on the symmetry of the initial and final SP
states, related hopping integrals are given by $\tau^{(21)}_{sd}$,
$\tau^{(21)}_{dd}$, $\tau^{(21)}_{pd}$, $\tau^{(21)}_{sp}$, and
$\tau^{(21)}_{pp}$. 

The analysis of Hamiltonian terms induced by the exchange interaction
can be performed along similar lines. 
We start with a necessary correction to the
Ising part of the exchange energy. When we have been constructing the
Schr\"odinger equation (\ref{schr1})-(\ref{schr3}) for
wavefunctions (\ref{wvfn}) of SPs, we have assumed that the Ising
contribution depends on the length of strings, but does not depend on
their geometry. That assumption is true for short strings, but for
longer strings there exists some exceptions from that rule. For
example, by inspecting the spin structure of the string state depicted
Fig. \ref{fig14} (d), we see that one broken  bond has
been saved for that state in comparison with strings which have the
same length and have the starting and ending points  which are not
NNs. That fact gives rise to the decrease by $J/2$ of the string-state
energy and to corrections to eigenenergies of $d$-wave and $p$-wave
SPs. Those corrections are given by $\iota^{(19)}_{dd}$ and
$\iota^{(19)}_{pp}$ respectively. Since the amendments are related to
strings of length at least 3 lattice spacings we neglect them in the
case of the $s$-wave SP because the weight of related string states in
the wavefunction of the s-wave SP is small.

\begin{figure}
\includegraphics[width=\columnwidth]{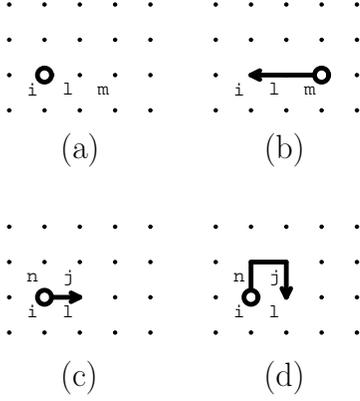}
\caption{\label{fig15} 
(a) and (b) A low-order process which involves the action of
the transverse part of the exchange term and gives rise to the hybridization
of SPs with different symmetry. (c) and (d) String states involved in a
process mediated by the XY part of the exchange term and resulting in
a correction to the eigenenergy of the $d$-wave SP.}
\end{figure}

Also the best known in literature process which determines to great
extent the overall shape of the single-particle energy-dispersion at
the low energy scale gives rise to the hybridization between SPs with
different symmetries. That process has been shown in
Fig. \ref{fig15} (a) and (b). 
Flipping spins at sites $l$ and $m$ in the string state depicted in
Fig. \ref{fig15} (a) by the transverse part of the exchange energy
gives rise to the string state depicted in Fig. \ref{fig15} (b) and
to the hybridization between SPs at sites $i$ and $m$. In that kind of
hybridization the $s$-wave state should be involved because the state
shown in Fig. \ref{fig15} (a) represents a bare hole created in the
N\'eel state. As we have shown before, such string states do not
contribute to wavefunctions of SPs with lower symmetry. The hopping
integrals which appear at terms coupling an $s$-wave SP with an
$s$-wave, a $d$-wave, and a $p$-wave SP located at a third NN site are
given by $\iota^{(3)}_{ss}$, $\iota^{(3)}_{ds}$, and
$\iota^{(3)}_{ps}$ respectively.

A similar mechanism gives rise to a correction to the eigenenergy of a
$d$-wave SP. The action of the XY term flips spins in the string state
depicted in Fig. \ref{fig15} (c) at sites $n$, $j$ and creates the
string state shown in Fig. \ref{fig15} (d). Due to the fact that
both string states contribute to the wavefunctions (\ref{wvfn}) of SPs
created at the same site $i$ and due to standard selection rules, the
contribution to ${\cal T}^{(oo')}_{\bf R}$ originating with the above
discussed process must be diagonal. For this reason, $p$-wave states
can not contribute to the related new matrix elements ${\cal
  T}^{(oo')}_{\bf R}$, because the process depicted in Fig.
\ref{fig15} (c), (d) involves string states which contribute to
different $p_x$-wave and $p_y$-wave SPs. Since the length of the
string depicted in Fig. \ref{fig15} (d) is 3 lattice spacings we
neglect the correction to the eigenenergy of the $s$-wave SP, because
we expect that it is small. The correction to the eigenenergy of the
$d$-wave SP is $-\iota^{(11)}_{dd}$. It also contains contributions
related to the coupling, in the same way, between longer-string states
obtained by letting holes at the ends of strings in Fig.
\ref{fig15} (c) and (d) to hop further along identical paths.

\begin{figure}
\includegraphics[width=\columnwidth]{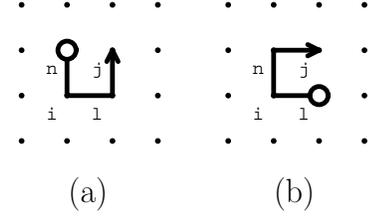}
\caption{\label{fig16} 
(a) and (b) Spuriously different string states which are
actually identical, what gives rise to the overlap between SPs created
at sites $n$ and $l$.}
\end{figure}

During the construction of SPs we have been assuming that
wavefunctions of SPs at different sites are orthogonal which turns out
not to be an exactly true assumption. On the other hand the overlap
between SPs is rather small because it originates with the overlap of
nominally different string states related with paths of relatively long
length. Such string states have been depicted in Fig. \ref{fig16}
(a) and (b). 
Both Figures \ref{fig16} (a) and (b) actually represent the same
state which consists of a hole at site $j$ and fluctuations at sites
$n$, $i$, $l$ created in the N\'eel state. The generated in this way
overlap between different SPs  also gives rise to a new contribution in the
Hamilton operator,
\begin{eqnarray}
&& p_yd,\hat{\bf x}+\hat{\bf y}, E_1^{\prime}
\omega^{(1)}_{pd}: \;C_{4v} \cdot H.C. \\&&
dd,\hat{\bf x}+ \hat{\bf y}, E_1^{\prime} \omega^{(1)}_{dd};\;
p_yp_x,\hat{\bf x}+ \hat{\bf y},E_1^{\prime} \omega^{(1)}_{pp}: \;
C_{4v}.
\end{eqnarray}

The analysis of other processes which give rise to new elements of
${\cal T}^{(oo')}_{\bf R}$ is rather straightforward and similar to the analysis of
previously discussed processes. Thus we do not discuss the remaining
contributions to ${\cal T}^{(oo')}_{\bf R}$ one by one.

Finally, we list here matrix matrix elements which determine the form of the
Hamilton matrix (\ref{hmatelksp}), (\ref{hmatel}),

\begin{eqnarray}
&&ss,0,E_1;\; dd,0,E^{\prime}_1;\; p_xp_x,0,E^{\prime}_1;\;
p_yp_y,0,E^{\prime}_1 \\ &&
ss,0,
\tau^{(6)}_{ss};\;
ss,\hat{\bf x}+\hat{\bf y},\tau^{(1)}_{ss};\; ss,2\hat{\bf x},
\chi^{(\beta)}_{ss;(2,0)}; \nonumber\\&& dd,0,  \tau^{(6)}_{dd};\; 
dd,2\hat{\bf x},\tau^{(14)}_{dd};\;p_x p_x, 0,\tau^{(6)}_{pp};\;   
\nonumber\\&&
p_xp_x,2\hat{\bf x}, \tau^{(14)}_{pp}:\;C_4
\\&&
ss,00,\tau^{(13)}_{ss};\;
ss,2\hat{\bf x},\iota^{(3)}_{ss};\;
ds,2\hat{\bf x},\chi^{(\gamma)}_{ds;(2,0)};\nonumber\\&& 
dd,0,\tau^{(13)}_{dd} ;\;
dd,2\hat{\bf x},\iota^{(7)}_{dd};\; 
dp_x,2\hat{\bf x},\iota^{(7)}_{dp};\;
\nonumber\\&&
p_x s,2\hat{\bf x},\chi^{(\gamma)}_{p_x s;(2,0)};\;  
p_x d,2\hat{\bf x},\chi^{(\gamma)}_{p_x d;(2,0)};\;\nonumber\\&& 
p_xp_x,0,\tau^{(13)}_{pp};\;
p_xp_x,2\hat{\bf x},\iota^{(7)}_{pp} :\;C_4 \cdot H.C.
\\&&
ss,\hat{\bf x}+\hat{\bf y},\iota^{(3)}_{ss};\; 
sd,\hat{\bf x}+\hat{\bf y},\tau^{(21)}_{sd};\;\nonumber\\&&
sp_x,\hat{\bf x}+\hat{\bf y},\tau^{(21)}_{sp};\;
dd,\hat{\bf x}+\hat{\bf y},\chi^{(\delta)}_{dd;(1,1)};\;\nonumber\\&& 
dd,0,-\iota^{(11)}_{dd};\; 
dd,2 \hat{\bf x},-\iota^{(7)}_{dd};\;\nonumber\\&&
d p_x,\hat{\bf x}+\hat{\bf y},\chi^{(\delta)}_{d p_x;(1,1)};\; 	
p_x d,\hat{\bf x}+\hat{\bf y},\chi^{(\delta)}_{p_x d;(1,1)};\;\nonumber\\&&
p_xd,2 \hat{\bf x},\iota^{(7)}_{pd};\; 
p_x p_x,\hat{\bf x}+\hat{\bf y},\chi^{(\delta)}_{p_x p_x;(1,1)};\;
\nonumber\\&&
p_y s,\hat{\bf x}+\hat{\bf y},\chi^{(\delta)}_{p_y s;(1,1)};\;
p_y d,\hat{\bf x}+\hat{\bf y},\chi^{(\delta)}_{p_y d;(1,1)};\; \nonumber\\&&
p_y p_x,\hat{\bf x}+\hat{\bf y},\chi^{(\delta)}_{p_y p_x;(1,1)} 
:\;C_{4v} \cdot H.C.
\\&&
ss,0,\chi^{(\epsilon)}_{ss;(0,0)};\;
ss,\hat{\bf x}+ \hat{\bf y},\tau^{(15)}_{ss};\; \nonumber\\&&
dd,0,\chi^{(\epsilon)}_{dd;(0,0)};\;
dd,\hat{\bf x}+ \hat{\bf y},\chi^{(\epsilon)}_{dd;(1,1)} ;\; \nonumber\\&&
p_x p_x,0,\chi^{(\epsilon)}_{p_x p_x;(0,0)} ;\;
p_x p_x,\hat{\bf x}+ \hat{\bf y},\tau^{(23)}_{p p} ;\;\nonumber\\&&
p_y p_x,\hat{\bf x}+ \hat{\bf y},\chi^{(\epsilon)}_{p_y p_x;(1,1)}:\; 
C_{4v}\\&&
p_yp_x,\hat{\bf x}+\hat{\bf y},\tau^{(21)}_{pp}:\; C_{2v} \cdot H.C.
\end{eqnarray}
and of the overlap matrix (\ref{ovmatelksp}),  (\ref{ovmatel}),
\begin{eqnarray}
&&p_yd,\hat{\bf x}+\hat{\bf y},\omega^{(1)}_{pd}:\;  C_{4v} \cdot H.C.
\\&&
dd,\hat{\bf x}+ \hat{\bf y},\omega^{(1)}_{dd} ;\;
p_yp_x,\hat{\bf x}+ \hat{\bf y},\omega^{(1)}_{pp}:\; C_{4v}.
\end{eqnarray}
where
\begin{eqnarray}
&& \chi^{(\beta)}_{ss;(2,0)}=\tau^{(2)}_{ss}+\tau^{(14)}_{ss} 
,\\&&
 \chi^{(\gamma)}_{ds;(2,0)}=\iota^{(3)}_{ds}+\tau^{(14)}_{ds} 
,\\&&
\chi^{(\gamma)}_{p_x s;(2,0)}=\iota^{(3)}_{ps}+\tau^{(14)}_{ps} 
,\\&&
\chi^{(\gamma)}_{p_x d;(2,0)}=\iota^{(7)}_{pd}+\tau^{(14)}_{pd} 
,\\&&
 \chi^{(\delta)}_{d d;(1,1)}=\iota^{(11)}_{dd}+\tau^{(17)}_{dd}+\tau^{(18)}_{dd}   
,\\&&
 \chi^{(\delta)}_{d p_x;(1,1)}=\iota^{(7)}_{pd}-\iota^{(11)}_{pd}+\tau^{(17)}_{pd}+\tau^{(18)}_{pd}   
,\\&&
 \chi^{(\delta)}_{p_x d;(1,1)}=\iota^{(7)}_{pd}+\iota^{(11)}_{pd}+\tau^{(23)}_{pd} 
,\\&&
 \chi^{(\delta)}_{p_x p_x;(1,1)}=\iota^{(7)}_{pp}+\iota^{(11)}_{pp} 
,\\&&
 \chi^{(\delta)}_{p_y s;(1,1)}=\iota^{(3)}_{ps}+\tau^{(15)}_{ps} 
,\\&&
 \chi^{(\delta)}_{p_y d;(1,1)}=\iota^{(7)}_{pd}+\tau^{(15)}_{pd}+\tau^{(17)}_{pd}+\tau^{(18)}_{pd} 
\nonumber\\
&& + (E_1^{\prime}-J/2)\omega^{(1)}_{pd} +\tau^{(21)}_{pd}    
,\\&&
 \chi^{(\delta)}_{p_y p_x;(1,1)}=\iota^{(7)}_{pp}+\tau^{(17)}_{pp}+\tau^{(18)}_{pp} 
,\\&&
 \chi^{(\epsilon)}_{ss;(0,0)}=\tau^{(5)}_{ss}+\tau^{(16)}_{ss} 
,\\&&
 \chi^{(\epsilon)}_{dd;(0,0)}=\tau^{(5)}_{dd}+\tau^{(16)}_{dd} +\iota^{(19)}_{dd}+\tau^{(22)}_{dd}
,\\&&
\chi^{(\epsilon)}_{dd;(1,1)}=\tau^{(15)}_{dd}
+(E_1^{\prime}-J/2)\omega^{(1)}_{dd} +\tau^{(21)}_{dd}\nonumber \\
&&-\tau^{(22)}_{dd}   
,\\&&
\chi^{(\epsilon)}_{p_x p_x;(0,0)}=\tau^{(16)}_{pp} +  \tau^{(19)}_{pp} 
,\\&&
\chi^{(\epsilon)}_{p_y p_x;(1,1)}=\tau^{(15)}_{pp} +(E_1^{\prime}-J/2)\omega^{(1)}_{pp}, 
\end{eqnarray}

\begin{eqnarray}
&&
\tau^{(1)}_{ss}=t^{\prime} \alpha_0^2 
;\\&&
\tau^{(2)}_{ss}=t^{\prime \prime} \alpha_0^2 
;\\&&
\iota^{(3)}_{ss} = \frac{J}{2} \sum_{\mu = 0} 3^{\mu} \alpha_{\mu} \alpha_{\mu+2}
, \;\nonumber \\&&
\iota^{(3)}_{ds} = \frac{J}{4} \sum_{\mu = 0} 3^{\mu} \alpha_{\mu} \alpha^{\prime}_{\mu+2}
, \nonumber \\&&
\iota^{(3)}_{ps} = -\frac{J}{2 \sqrt{2}} \sum_{\mu = 0} 3^{\mu} \alpha_{\mu} \alpha^{\prime}_{\mu+2}
;\\&&
\tau^{(5)}_{ss}=t^{\prime} \alpha_1^2 
, \;
\tau^{(5)}_{dd}=-\frac{t^{\prime}}{4} (\alpha^{\prime}_1)^2
;\\&&
\tau^{(6)}_{ss}=t^{\prime \prime} \alpha_1^2 
, \;
\tau^{(6)}_{dd}=\frac{t^{\prime \prime}}{4} (\alpha^{\prime}_1)^2 
, \;\nonumber\\&&
\tau^{(6)}_{pp}=-\frac{t^{\prime \prime}}{2} (\alpha^{\prime}_1)^2 
;\\&&
\iota^{(7)}_{dd} = \frac{J}{8} \sum_{\mu = 1} 3^{\mu-1} \alpha^{\prime}_{\mu} \alpha^{\prime}_{\mu+2}
, \;\nonumber\\&&
\iota^{(7)}_{pd} = -\frac{J}{4\sqrt{2}} 
\sum_{\mu = 1} 3^{\mu-1} \alpha^{\prime}_{\mu} \alpha^{\prime}_{\mu+2}
, \nonumber \\&&
\iota^{(7)}_{pp} = \frac{J}{4} \sum_{\mu = 1} 3^{\mu-1} \alpha^{\prime}_{\mu} \alpha^{\prime}_{\mu+2}
;\\&&
\iota^{(11)}_{dd} = \frac{J}{8}[\alpha^{\prime}_1\alpha^{\prime}_3+
 2 \sum_{\mu = 2} 3^{\mu-2} \alpha^{\prime}_{\mu} \alpha^{\prime}_{\mu+2}]
, \;\nonumber\\&&
\iota^{(11)}_{pd} = -\frac{J}{4 \sqrt{2}}[\alpha^{\prime}_1\alpha^{\prime}_3+
 2 \sum_{\mu = 2} 3^{\mu-2} \alpha^{\prime}_{\mu} \alpha^{\prime}_{\mu+2}]
, \nonumber \\&&
\iota^{(11)}_{pp} = -\frac{J}{4}[\alpha^{\prime}_1\alpha^{\prime}_3+
 2 \sum_{\mu = 2} 3^{\mu-2} \alpha^{\prime}_{\mu} \alpha^{\prime}_{\mu+2}]
;\\&&
\tau^{(13)}_{ss} = 2 t^{\prime} \sum_{\mu = 2} 3^{\mu-2} \alpha_{\mu}^2
, \;
\tau^{(13)}_{dd} = \frac{t^{\prime}}{2} \sum_{\mu = 2} 3^{\mu-2} (\alpha^{\prime}_{\mu})^2
, \nonumber \\&&
\tau^{(13)}_{pp} = t^{\prime} \sum_{\mu = 2} 3^{\mu-2} (\alpha^{\prime}_{\mu})^2
;\\&&
\tau^{(14)}_{ss}=t^{\prime \prime} \alpha_2^2 
, \;
\tau^{(14)}_{ds}=\frac{t^{\prime \prime}}{2} \alpha_2 \alpha^{\prime}_2
, \;\nonumber\\&&
\tau^{(14)}_{ps}=-\frac{t^{\prime \prime}}{\sqrt{2}} \alpha_2 \alpha^{\prime}_2
, 
\tau^{(14)}_{dd}=\frac{t^{\prime \prime}}{4} (\alpha^{\prime}_2)^2
, \; \nonumber\\&&
\tau^{(14)}_{pd}=-\frac{t^{\prime \prime}}{2\sqrt{2}} (\alpha^{\prime}_2)^2
, \;
\tau^{(14)}_{pp}=-\frac{t^{\prime \prime}}{2} (\alpha^{\prime}_2)^2
;\\&&
\tau^{(15)}_{ss}=t^{\prime } \alpha_2^2 
, \;
\tau^{(15)}_{ps}=-\frac{t^{\prime}}{\sqrt{2}} \alpha_2 \alpha^{\prime}_2
, \; \nonumber\\&&
\tau^{(15)}_{dd}=-\frac{t^{\prime \prime}}{4} (\alpha^{\prime}_2)^2
, 
\tau^{(15)}_{pd}=-\frac{t^{\prime}}{2\sqrt{2}} (\alpha^{\prime}_2)^2
, \;\nonumber\\&&
\tau^{(15)}_{pp}=-\frac{t^{\prime}}{2} (\alpha^{\prime}_2)^2
;\\&&
\tau^{(16)}_{ss}=t^{\prime \prime} \sum_{\mu=2}3^{\mu-2}\alpha_\mu^2 
, \;
\tau^{(16)}_{dd}=\frac{t^{\prime \prime}}{4} \sum_{\mu=2}3^{\mu-2}(\alpha^{\prime}_\mu)^2 
, \nonumber \\&&
\tau^{(16)}_{pp}=\frac{t^{\prime \prime}}{2} \sum_{\mu=2}3^{\mu-2}(\alpha^{\prime}_\mu)^2 
;\\&&
\tau^{(17)}_{dd}=-\frac{t^{\prime \prime }}{4} (\alpha^{\prime}_3)^2 
, \; 
\tau^{(17)}_{pd}=-\frac{t^{\prime \prime }}{2 \sqrt{2}} (\alpha^{\prime}_3)^2 
, \; \nonumber\\&&
\tau^{(17)}_{pp}=-\frac{t^{\prime \prime }}{2} (\alpha^{\prime}_3)^2
;\\&&
\tau^{(18)}_{dd}=-\frac{t^{\prime}}{4} (\alpha^{\prime}_3)^2 
, \; 
\tau^{(18)}_{pd}=-\frac{t^{\prime}}{2 \sqrt{2}} (\alpha^{\prime}_3)^2 
, \; \nonumber\\&&
\tau^{(18)}_{pp}=-\frac{t^{\prime}}{2} (\alpha^{\prime}_3)^2 
;\\&&
\iota^{(19)}_{dd} = -\frac{J}{8}[(\alpha^{\prime}_3)^2+
 2 \sum_{\mu = 4} 3^{\mu-4} (\alpha^{\prime}_{\mu})^2]
, \;\nonumber\\&&
\iota^{(19)}_{pp} = -\frac{J}{4}[(\alpha^{\prime}_3)^2+
 2 \sum_{\mu = 4} 3^{\mu-4} (\alpha^{\prime}_{\mu})^2]
;\\&&
\tau^{(21)}_{sd}=\frac{t}{2} \alpha^{\prime}_3 \alpha_2 
, \; 
\tau^{(21)}_{dd}=-\frac{t}{4} \alpha^{\prime}_3 \alpha^{\prime}_2 
, \; \nonumber\\&&
\tau^{(21)}_{pd}=-\frac{t}{2 \sqrt{2}} \alpha^{\prime}_3 \alpha^{\prime}_2 
, \;
\tau^{(21)}_{sp}=\frac{t}{\sqrt{2}} \alpha^{\prime}_3 \alpha_2 
, \; \nonumber \\&&
\tau^{(21)}_{pp}=-\frac{t}{2} \alpha^{\prime}_3 \alpha^{\prime}_2 
;\\&&
\tau^{(22)}_{dd}=-\frac{t^{\prime}}{4} (\alpha^{\prime}_3)^2
;\\&&
\tau^{(23)}_{pd}=-\frac{t^{\prime}}{2 \sqrt{2}} (\alpha^{\prime}_3)^2 , \; 
\tau^{(23)}_{pp}=-\frac{t^{\prime}}{2 } (\alpha^{\prime}_3)^2 
\end{eqnarray}
and
\begin{eqnarray}
&&
\omega^{(1)}_{dd}=-\frac{(\alpha^{\prime}_3)^2+
 2 \sum_{\mu = 4} 3^{\mu-4} (\alpha^{\prime}_{\mu})^2}{4}
, \;\nonumber \\&&
\omega^{(1)}_{pd}=-\frac{(\alpha^{\prime}_3)^2+
 2 \sum_{\mu = 4} 3^{\mu-4} (\alpha^{\prime}_{\mu})^2}{2\sqrt{2}}
, \nonumber \\&&
\omega^{(1)}_{pp}=-\frac{(\alpha^{\prime}_3)^2+
 2 \sum_{\mu = 4} 3^{\mu-4} (\alpha^{\prime}_{\mu})^2}{2}.
\end{eqnarray}

\section{Optical conductivity of doped antiferromagnets}
\label{appb}
The optical spectrum evaluated by us is determined by following
contributions to matrix elements $\delta^{(n)} \langle \Psi^{(o^{\prime},0)}_{{\bf R}_i}|j_x
|\Psi^{(o),0}_{\bf R=0}\rangle \equiv s^{(n)}_{o^{\prime},o}({\bf R}_i)$
where $n$ labels different contributions,
\begin{eqnarray}
&&
s^{(1)}_{p_x,s}({\bf 0})= - 2 i t \sum_{\mu=0} \alpha_{\mu} 
\alpha^{\prime}_{\mu+1}/\sqrt{2}
, \;\nonumber\\
&&  s^{(1)}_{s,p_x}({\bf 0})= - 2 i t \sum_{\mu=1} \alpha^{\prime}_{\mu} 
\alpha_{\mu+1}/\sqrt{2}, \nonumber\\
&&   s^{(1)}_{d,p_x}({\bf 0})= - 2 i t \sum_{\mu=1}
\alpha^{\prime}_{\mu}
\alpha^{\prime}_{\mu+1}/(2 \sqrt{2}), \;\nonumber\\
&&  s^{(1)}_{p_x,d}({\bf
  0})=s^{(1)}_{d,p_x}({\bf 0}) \\
&&
s^{(2)}_{p_x,s}({\bf 0})=  2 i t \sum_{\mu=2} \alpha_{\mu} 
\alpha^{\prime}_{\mu-1}/\sqrt{2}
, \;\nonumber\\
&&  s^{(2)}_{s,p_x}({\bf 0})=  2 i t \sum_{\mu=1} \alpha^{\prime}_{\mu} 
\alpha_{\mu-1}/\sqrt{2}, \nonumber\\
&&   s^{(2)}_{d,p_x}({\bf 0})= -2 i t \sum_{\mu=2}
\alpha^{\prime}_{\mu}
\alpha^{\prime}_{\mu-1}/(2 \sqrt{2}), \;\nonumber\\
&&  s^{(2)}_{p_x,d}({\bf
  0})=s^{(2)}_{d,p_x}({\bf 0}) \\
&&
s^{(3)}_{s,s}(\hat{\bf x}\pm  \hat{\bf y})=-i t^{\prime} \alpha_{0}^2 \\
&&
s^{(4)}_{s,s}(-\hat{\bf x}\pm \hat{\bf y})=-s^{(3)}_{s,s}(\hat{\bf x}
+ \hat{\bf y}) \\
&&
s^{(5)}_{s,p_x}({\bf 0})= 4 i t^{\prime} \alpha^{\prime}_{1} \alpha_{1}/\sqrt{2} 
, \; \nonumber \\&&
 s^{(5)}_{d,p_x}({\bf 0})= - 4 i t^{\prime} (\alpha^{\prime}_{1})^2 /(2 \sqrt{2}) \\
&&
s^{(6)}_{p_x,s}({\bf 0})=- s^{(5)}_{s,p_x}({\bf 0})
, \; \nonumber \\&& s^{(6)}_{p_x,d}({\bf 0})= s^{(5)}_{d,p_x}({\bf 0}) \\
&&
s^{(7)}_{s,s}({2 \hat{\bf x} })=-2 i t^{\prime \prime} \alpha_{0}^2
,\; \nonumber \\&&  s^{(7)}_{s,s}({-2 \hat{\bf x} })= -s^{(7)}_{s,s}({2 \hat{\bf x} })\\
&&   s^{(8)}_{p_x,s}({\bf 0})= -4 i t^{\prime \prime} \alpha_{1}
\alpha^{\prime}_{1}/\sqrt{2}, \; \nonumber \\&& 
s^{(8)}_{s,p_x}({\bf 0})= -s^{(8)}_{p_x,s}({\bf 0}), \nonumber  \\
&&   s^{(8)}_{d,p_x}({\bf 0})= 4 i t^{\prime \prime} (\alpha^{\prime}_{1})^2/(2 \sqrt{2}), \; \nonumber \\&& 
s^{(8)}_{p_x,d}({\bf 0})= -s^{(8)}_{d,p_x}({\bf 0}).
\end{eqnarray}
For example contributions No. $1$ and $2$ are related to shortening
and elongating strings by the current operator, Fig. \ref{fig11}, while
contributions No. $3$ and $4$ to shifts between next NN 
sites, Fig. \ref{fig13}(a) and (b). We have considered processes
involving strings of length up to two for matrix elements of the
$t$-term in the current operator and up to one for matrix elements of
$t^\prime$ and $t^{\prime \prime}$ terms.

\end{appendix}

\end{document}